\documentclass[twocolumn,10pt]{article}
\usepackage[a4paper,margin=1.9cm,columnsep=0.7cm]{geometry}

\usepackage[T1]{fontenc}
\usepackage[utf8]{inputenc}
\usepackage{etoolbox}
\usepackage{xspace}
\usepackage[numbers]{natbib}
\usepackage{amsmath,amssymb,amsfonts}
\usepackage{graphicx}
\usepackage{textcomp}
\usepackage{algorithm}
\usepackage{algorithmic}
\usepackage{booktabs,array,multirow,makecell}
\usepackage{stfloats}
\usepackage{caption}
\usepackage{float}
\usepackage{xcolor}
\usepackage{hyperref}
\usepackage{url}

\hypersetup{colorlinks=true,linkcolor=black,citecolor=black,urlcolor=blue}

\newcommand{\credit}[1]{}

\newcommand{\methodname}{MOSAIC}

\begin{document}

\twocolumn[{%
  \begin{center}
    {\LARGE\bfseries MOSAIC: Modality-agnostic Spectral Alignment for
     Federated Image-level Weakly Supervised Tumor Segmentation under
     Client-specific Missing Modalities\par}
    \vspace{1.2em}
    {\large
      Tarun Kumar Garg\textsuperscript{1}\quad
      Vaanathi Sundaresan\textsuperscript{2,\,$\dagger$}\par}
    \vspace{0.8em}
    {\normalsize
      \textsuperscript{1}Department of IISc Mathematics Initiative, Indian Institute of Science, Bengaluru 560012, Karnataka, India\par
      \textsuperscript{2}Department of Computational and Data Sciences, Indian Institute of Science, Bengaluru 560012, Karnataka, India\par}
    \vspace{0.5em}
    {\footnotesize
      \textsuperscript{$\dagger$}Corresponding author: \href{mailto:vaanathi@iisc.ac.in}{vaanathi@iisc.ac.in}\par}
  \end{center}
  \vspace{1em}
  \begin{center}
  \begin{minipage}{0.92\textwidth}
    \noindent\textbf{Abstract.}\;
Trustworthy multimodal fusion in clinical settings requires handling incomplete and heterogeneous modality subsets across institutions, where privacy constraints prohibit centralized data sharing. Federated learning (FL) mitigates data-sharing constraints but suffers from client-specific missing modalities, where institutions possess incomplete multimodal subsets, degrading fusion quality and segmentation performance. While FL and weak supervision have been studied separately, their joint use with image-level labels under heterogeneous missing modalities remains unaddressed. We propose \textbf{MOSAIC}, the first modality-agnostic federated framework for weakly supervised binary tumor segmentation under client-specific missing modalities. We introduce a client-specific modality-alignment module that fuses available channels into a shared latent space without prior knowledge of modality identity, a spectral prototype alignment loss that reconciles cross-client distribution shift using compact non-invertible frequency-domain statistics, and a dedicated federated refinement network that denoises the resulting CAM pseudo-labels into accurate masks, breaking the accuracy ceiling of weak supervision. Experiments on three multi-institutional brain tumor benchmarks (FeTS2022, BraTS-MEN, and BraTS-SSA) demonstrate significant improvements over all image, box, and point-supervised baselines, approaching fully supervised accuracy using only image-level labels and reaching $0.84$ Dice on FeTS2022. Dynamic new client addition enables previously unseen institutions to join an already-trained federation within 0.01--0.04 Dice without retraining. Code is available at \url{https://github.com/Tarun2201/MOSAIC}.\par
    \vspace{0.6em}
  \end{minipage}
  \end{center}
  \vspace{1.5em}
}]

\section{Introduction}
\label{sec:introduction}
Trustworthy multimodal fusion of medical imaging data is a prerequisite for reliable diagnosis, treatment planning, and longitudinal monitoring of disease progression~\cite{zhang2022mmformer, hayat2022medfuse}. In neuro-oncology, fusing complementary information across multi-parametric MRI sequences, each capturing distinct tissue contrasts and pathological signatures, provides the volumetric and boundary measurements that guide surgical resection, radiotherapy target definition, and response assessment. Deep neural networks achieve strong performance when trained on large annotated cohorts, yet their clinical deployment is constrained by two persistent barriers: expert annotations are scarce and expensive, and patient data are siloed across institutions by privacy regulation.

Dense voxel-level annotation demands scarce neuroradiologist time, is subject to substantial inter-observer variability, and cannot be assembled at scale. Weakly supervised segmentation (WSS) offers a pragmatic alternative by learning from coarse labels that are far cheaper to obtain. Image-level labels, indicating only whether a structure is present, are the most economical, often extractable directly from existing radiology reports. Class activation maps (CAMs) and their variants recover spatial localization from such labels \cite{selvaraju2017_gradcam, chen2024_wsss}, but the resulting pseudo-labels are coarse and noisy, making accurate boundary delineation a central open challenge.

Privacy regulation and institutional governance prohibit centralized aggregation of medical images. Federated learning (FL) resolves this by enabling hospitals to jointly train a shared model while data never leave the local site \cite{sheller2020federated}. Standard FL optimizers address client drift~\cite{li2020federated}, feature distribution shift~\cite{li2021fedbn}, and inter-client alignment~\cite{li2021moon}, but universally assume dense supervision and complete modality availability. Crucially, FL alleviates the data-sharing barrier but not the annotation barrier, making the combination of federated learning with weak supervision both natural and clinically compelling.

This combination is further complicated by modality heterogeneity specific to multi-institutional imaging: acquisition protocols differ across sites, so clients routinely possess different subsets of MRI modalities, rendering standard multimodal fusion architectures inapplicable. When modalities are missing, fusion quality degrades non-uniformly across clients: modality-incomplete clients generate spurious false positives from absent tissue contrast, and CAM pseudo-labels become unreliable without the discriminative channels that complete multimodal fusion would otherwise provide. Methods designed for missing modalities in FL rely on fully supervised training rather than image-level labels and modality-specific encoders that assume explicit knowledge of each client's modality set \cite{bernecker2022_fednorm, peng2024_fedmm, liu2025_fedmepd}. The few existing federated WSS methods \cite{feddm2023tmi, lin2023fedicra, lin2025fedlppa} do not address missing modalities, and no existing work addresses all three challenges jointly. 

To address this gap, we propose \textbf{MOSAIC}, the first modality-agnostic FL framework for weakly supervised binary tumor segmentation under client-specific missing modalities, and the first to jointly address federated learning, image-level weak supervision, and heterogeneous multimodal missing modalities within a single end-to-end architecture. A lightweight client-specific alignment module maps each client's available modalities into a shared latent space using only channel count, enabling modality-agnostic multimodal fusion without metadata or bookkeeping. A spectral prototype alignment (SPA) loss aligns clients via compact frequency-domain band energies rather than raw spatial features, preserving privacy while measurably reducing cross-client feature divergence. CAM pseudo-labels produced by this aligned model are refined by a dedicated federated segmentation network, breaking the accuracy ceiling of noisy weak supervision. 
Our main contributions are as follows:
\begin{itemize}

    \item \textbf{First federated framework for image-level WSS under missing modalities.} We present the first modality-agnostic federated framework for end-to-end weakly supervised tumor segmentation under client-specific missing modalities, evaluated against sixteen baselines spanning four supervision levels across three clinically distinct datasets, including FeTS2022, BraTS-MEN, and BraTS-SSA, consistently outperforming all prior methods.

    \item \textbf{Modality-agnostic client alignment.} We introduce a 
    client-specific alignment module conditioned solely on available channel count, eliminating modality bookkeeping and enabling modality-agnostic multimodal fusion under privacy constraints, allowing clients with unseen modality combinations to join an already-trained federation without retraining.

    \item \textbf{Spectral prototype alignment with empirical divergence validation.} We propose a privacy-preserving SPA loss that aligns clients via compact frequency-domain band energies, validated empirically using MMD$^2$ and Fr\'{e}chet distance, demonstrating measurable cross-client feature divergence reduction at two network depths.

    \item \textbf{Two-phase weakly supervised refinement pipeline.} We 
    propose a two-phase architecture in which federated CAM pseudo-labels are progressively refined by a dedicated segmentation network, with each component validated via leave-one-out ablation.

    \item \textbf{Comprehensive evaluation under realistic federation 
    conditions.} We conduct systematic experiments covering modality 
    assignment sensitivity, dynamic new client addition, scalability under increasing client counts, SPA spectral resolution ablation over $\Omega \in \{2,4,6,8,10,12,14,16\}$ radial frequency bands, and aleatoric uncertainty quantification across all three datasets.
\end{itemize}

\section{Related Work}
\label{sec:related}
Our work bridges five research areas: weakly supervised segmentation, federated learning for medical imaging, federated weakly supervised segmentation, missing-modality learning and multimodal fusion, and frequency-domain alignment in federated learning.

\subsection{Weakly supervised segmentation}
\label{subsec:rw_wss}
%Weakly supervised segmentation (WSS) recovers pixel-level masks from inexpensive annotations, with image-level labels being the most economical. 
Class activation mapping (CAM) established the 
dominant paradigm for recovering spatial localization from image-level 
labels. Subsequent methods improved localization through gradient-based 
weighting~\cite{selvaraju2017_gradcam}, 
forward-pass scoring~\cite{score_cam}, multi-level 
aggregation~\cite{jiang2021layercam}, affinity-based pseudo-label 
propagation~\cite{ahn2019irnet}, and self-supervised equivariant 
attention~\cite{wang2020seam}. Erasing-based strategies further improved activation coverage by iteratively suppressing discriminative regions to force the network to discover complementary cues. More recently, transformer-based methods have leveraged self-attention to produce more complete and spatially consistent activation maps \cite{ xu2022multi}. In medical imaging, WSS has been applied to polyp segmentation, skin lesion detection \cite{qin2023portable}, and brain tumor localization~\cite{ame_cam}, where image-level labels are particularly economical as they can be derived directly from radiology reports. However, all existing WSS methods assume centralized data and homogeneous modality availability, limiting their applicability in multi-institutional settings. To our knowledge, MOSAIC is the first to extend image-level WSS to a federated setting with client-specific missing modalities.

\subsection{Federated learning for medical image segmentation}
\label{subsec:rw_fl}
FedAvg~\cite{mcmahan2017communication} introduced collaborative training without raw data sharing. Subsequent work addressed client drift (FedProx~\cite{li2020federated}), feature distribution shift 
(FedBN~\cite{li2021fedbn}), and inter-client contrastive alignment (MOON~\cite{li2021moon}). Personalized FL 
methods~\cite{fallah2020personalized} introduced client-specific components alongside a shared global backbone. Despite their effectiveness, all these methods assume complete and identical modality availability across clients, an assumption routinely violated in multi-centre deployments where acquisition protocols and scanner availability 
differ across institutions.

\subsection{Federated weakly supervised segmentation}
\label{subsec:rw_fedwss}
FedDM~\cite{feddm2023tmi} calibrates noisy pseudo-labels generated from bounding-box-level supervision while mitigating client drift using hierarchical gradient de-conflicting. FedICRA~\cite{lin2023fedicra} unified heterogeneous weak supervision through site-contrastive representation learning, while FedLPPA~\cite{lin2025fedlppa}, a recent state of the art method leverages personalized prompts and adaptive decoder aggregation. FL-W3S~\cite{madni2025fl} explored federated weakly supervised semantic segmentation using image-level labels for white blood cell segmentation. These methods use denser form of supervision (e.g., boxes, points, scribbles, and blocks) and none of them address client-specific missing modalities alongside weak supervision, leaving the joint setting unaddressed.

\subsection{Missing-modality learning and multimodal fusion}
\label{subsec:rw_missing}
Robustness to missing modalities has been studied extensively in 
centralized settings. Shared latent projection methods map each 
available modality into a common representation and fuse any available 
subset~\cite{ding2021rfnet}. Knowledge distillation 
approaches transfer representations from a full-modality teacher to a 
modality-incomplete student~\cite{hu2020kdnet}. Generative methods reconstruct missing modalities from available ones, while transformer-based architectures have addressed missing modality fusion through masked reconstruction~\cite{zhang2022mmformer}. Beyond medical imaging, uncertainty-aware fusion frameworks quantify modality-level confidence under incomplete observations using evidential deep learning~\cite{xie2023exploring}, and dynamic fusion mechanisms adaptively weight modality contributions based on input completeness \cite{ding2021rfnet}. Multimodal fusion under asynchronous and misaligned modalities has been addressed in clinical prediction using attention-based temporal alignment. Multimodal fusion has also been explored in clinical prediction by integrating heterogeneous sources such as clinical time-series and medical imaging~\cite{hayat2022medfuse}. These approaches address incomplete and asynchronously observed modalities by selectively fusing available information. In federated settings, existing missing modality methods rely on modality-specific encoders~\cite{liu2025_fedmepd}, prototype-guided aggregation, per-modality feature 
extractors, or modality-aware 
normalisation~\cite{bernecker2022_fednorm}, and all require full 
pixel-level supervision. In practice, modality availability is 
determined by scanner access, patient compliance, and institutional 
procurement, making client-specific missing modalities the norm 
rather than the exception in real multi-centre deployments. In 
contrast, MOSAIC operates under image-level weak supervision and 
conditions only on available channel count, enabling truly 
modality-agnostic multimodal fusion without pixel-wise annotation or 
modality metadata.

\subsection{Frequency-Domain Alignment in Federated Learning}
\label{subsec:rw_freq}
The use of frequency-domain representations for domain alignment has 
roots in classic signal processing, where amplitude and phase 
decomposition separates style from content in natural 
images \cite{yang2020fda}. In deep learning, style transfer using Gram matrices of frequency statistics \cite{gatys2016image} demonstrated that spectral representations capture distributional 
properties efficiently. Applied to federated heterogeneity, HarmoFL~\cite{jiang2022harmofl} harmonized Fourier-domain amplitudes 
to mitigate client drift, and FedDG~\cite{liu2021feddg} performed 
federated domain generalization via continuous frequency space 
interpolation. These ideas have been extended to polyp 
segmentation~\cite{pan2025fdgpolyp} and heterogeneous medical 
imaging~\cite{wang2025fedfat}. Prototype exchange has further proven 
effective for privacy-preserving cross-client alignment without 
exposing spatial features~\cite{tan2022fedproto}. Existing 
frequency-domain methods assume complete modality availability. MOSAIC 
extends this line of work to the missing-modality setting by 
constructing spectral prototypes from whichever modalities a client 
holds, aligning only compact band energies across clients while 
preserving multimodal fusion quality under heterogeneous inputs.

\subsection{Privacy-Preserving Federated Fusion}
Privacy constraints fundamentally shape what information can be shared during multimodal fusion across institutions. Differential privacy mechanisms \cite{mcmahan2017learning} add calibrated noise to model updates to bound information leakage, while secure aggregation protocols \cite{bonawitz2017practical} encrypt client contributions before server aggregation. Recent work has shown that sharing compact statistics rather than raw features or gradients provides an effective privacy-utility trade-off in federated settings~\cite{tan2022fedproto,wang2025fedfat}. MOSAIC adopts this principle by sharing only non-invertible frequency-domain band energies rather than spatial feature maps, ensuring that no raw imaging data or reconstructible representation ever leaves a client.

In summary, no existing method unifies federated learning, image-level weak supervision, and client-specific missing modalities; MOSAIC directly targets this gap by enabling modality-agnostic multimodal fusion across heterogeneous institutions without annotations beyond image-level labels.

\section{Methodology}
\label{sec:method}

\subsection{Problem Formulation}
\label{subsec:problem}
Let $\mathcal{M} = \{m_1, \ldots, m_{|\mathcal{M}|}\}$ denote the set 
of imaging modalities present across the federation, and let a full-modal slice be $X^{\mathcal{M}} \in \mathbb{R}^{|\mathcal{M}| \times H \times W}$, with $H$ and $W$ the spatial dimensions. We consider a federation of $K$ clients (institutions), where each client $k$ holds a private dataset:
\begin{equation}
    \mathcal{S}_k = \left\{(X^k_i, y^k_i)\right\}_{i=1}^{N_k}, \quad 
    X^k_i \in \mathbb{R}^{|\mathcal{M}_k| \times H \times W},
\end{equation}
with a client-specific modality subset $\mathcal{M}_k \subseteq 
\mathcal{M}$ and a binary image-level label $y^k_i \in \{0,1\}$ 
indicating tumor presence. The subsets are heterogeneous: no modality is guaranteed to be present at every client, and some clients hold only a single modality. This induces two coupled challenges: cross-client distribution shift from heterogeneous multimodal inputs, and unreliable pseudo-labels from incomplete modality fusion, that standard federated methods cannot resolve jointly.

\begin{figure*}
    \centering
    \includegraphics[width=0.95\linewidth]{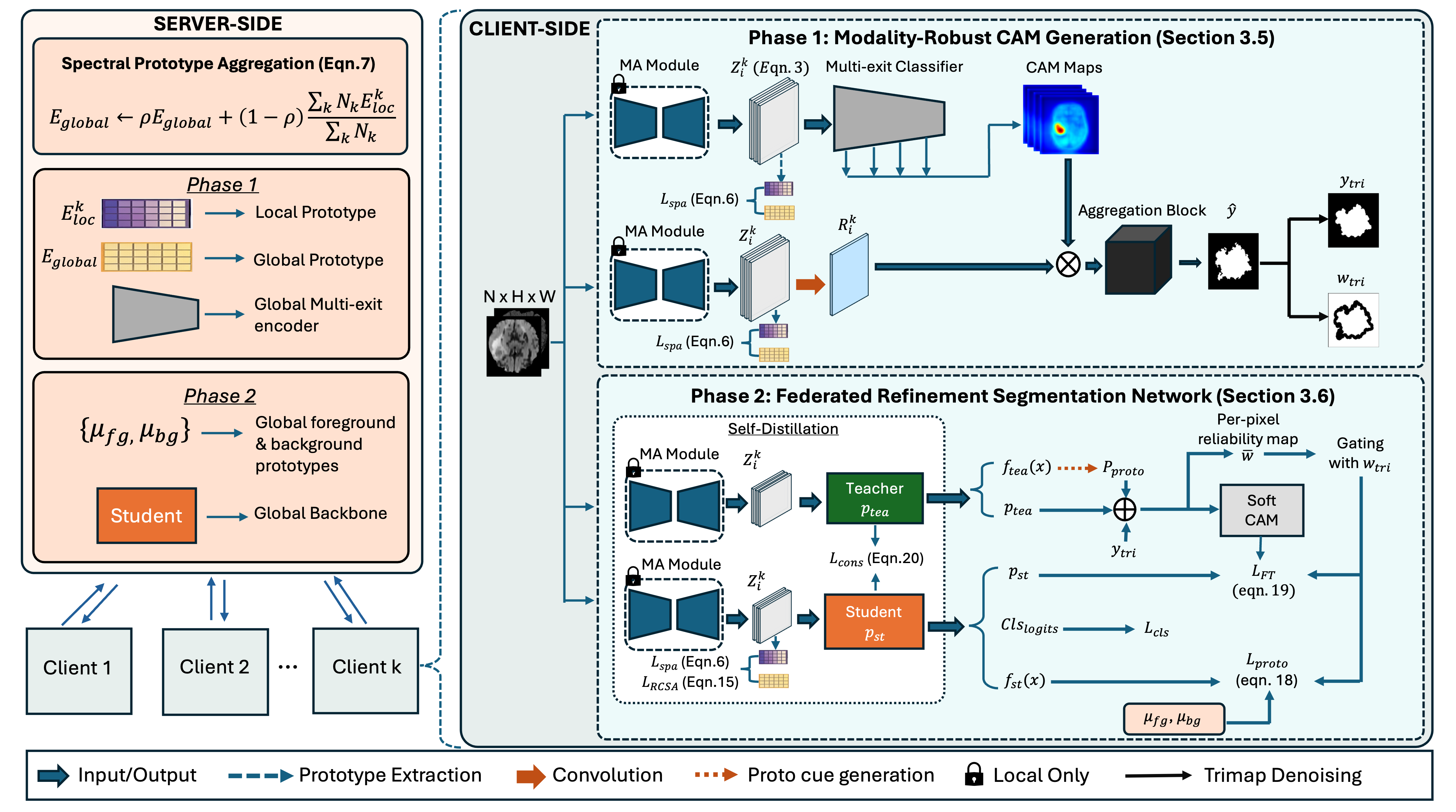}
    \caption{Overview of \methodname{} comprising $K$ clients and a central server. \textbf{Phase~1:} Each client passes its available modalities through a local modality-alignment (MA) module to obtain $Z_{i}^k$. A multi-exit classifier generates CAMs which, together with a projection of $Z_{i}^k$, are processed by an aggregation network to produce pseudo-labels $\hat{y}$. The SPA loss aligns local prototypes $E_{\mathrm{loc}}$ to the global prototype $E_{\mathrm{global}}$. \textbf{Phase~2:} An EMA teacher--student framework combines the teacher prediction $p_{\mathrm{tea}}$, prototype cue $p_{\mathrm{proto}}$, and trimap target $y_{\mathrm{tri}}$ into a soft CAM that supervises the student via $\mathcal{L}_{\mathrm{FT}}$. The prototype-alignment loss $\mathcal{L}_{\mathrm{proto}}$ aligns student features $f_{\mathrm{st}}$ to federated prototypes $(\mu_{\mathrm{fg}},\mu_{\mathrm{bg}})$.}
    \label{fig:framework}
\end{figure*}
Our objective is to obtain client-personalized models $\theta_k$ that 
map a modality-incomplete input to a spatial tumor-probability map 
$P_{\theta_k}(X^k_i) \in [0,1]^{H \times W}$ using only image-level 
supervision, while data never leave the local site and only model 
parameters and compact statistics are communicated. Denoting by 
$\mathcal{L}_k$ the local objective at client $k$ and $N = \sum_k 
N_k$, federated training minimizes the size-weighted sum of local 
objectives
\begin{equation}
    \min_{\{\theta_k\}} \sum_{k=1}^{K} \frac{N_k}{N} \mathcal{L}_k
    \left(\mathcal{S}_k; \theta_k\right).
\end{equation}

\subsection{Framework Overview}
\label{subsec:overview}
MOSAIC operates in two federated phases that share a common 
modality-agnostic core as shown in Figure~\ref{fig:framework}. \textit{Phase~1} (modality-robust CAM generation) trains a weakly supervised classifier and aggregation network to produce binary tumor pseudo-masks from image-level labels, using a classification-and-aggregation pipeline~\cite{dhamale2025_interclass}. \textit{Phase~2} (federated refinement segmentation) trains a dedicated segmentation network on these pseudo-masks, progressively denoising noisy weak supervision into accurate binary predictions.

Both phases are made robust to missing modalities through two shared components: a \textit{client-specific modality-alignment module} (Sec.~\ref{subsec:modality}) that fuses each client's heterogeneous modality subset into a shared representation, and a \textit{spectral prototype alignment} (SPA) loss (Sec.~\ref{subsec:spa}) that reconciles residual cross-client distribution shift by exchanging only compact, non-invertible frequency statistics. Together, these components implement a privacy-preserving multimodal fusion strategy that requires no prior knowledge of modality identity and accommodates clients with previously unseen modality combinations. Phase~1 extends the weakly supervised classification-and-aggregation pipeline of~\cite{dhamale2025_interclass}, originally designed for centralized homogeneous-modality settings, to the federated missing-modality regime by integrating the modality-alignment module and SPA loss, enabling reliable CAM pseudo-label generation across clients with heterogeneous modality subsets. Phase~2 then addresses the pseudo-label noise amplified by modality incompleteness through a dedicated federated refinement network, breaking the accuracy ceiling of the original pipeline. Phase~1 is optimized with FedProx and Phase~2 with a reliability-weighted FedAvg.

\subsection{Client-Specific Modality Alignment}
\label{subsec:modality}
Heterogeneous modality subsets prevent the shared, modality-dependent early layers from being aggregated directly without destroying complementary cross-modal information. Naive concatenation or zero-imputation of missing channels is inadequate because it either fixes the input dimensionality (preventing unseen modality combinations) or introduces spurious signals that corrupt the fused representation. We therefore equip each client $k$ with a lightweight alignment module $A_{\phi_k}$ that implements a learned multimodal fusion operation mapping its local input into a common latent 
representation of fixed dimensionality,
\begin{equation}
    Z^k_i = A_{\phi_k}(X^k_i), \quad Z^k_i \in \mathbb{R}^{D \times 
    H \times W}.
    \label{eq:modality_align}
\end{equation}
The module is a shallow U-Net~\cite{ronneberger2015u} with instance 
normalization that projects the $|\mathcal{M}_k|$ available channels 
onto $D$ canonical channels. Crucially, $A_{\phi_k}$ requires no 
prior knowledge of modality types and depends only on the number of 
channels a client holds, so clients with previously unseen modality 
combinations can join the federation without any modality bookkeeping. 
The parameters $\phi_k$ are kept local and are never aggregated, which 
isolates modality-dependent fusion preprocessing at each site while 
allowing the shared downstream network to remain modality-agnostic. 
This design implements a form of personalized federated 
fusion where the fusion front-end is client-specific while the semantic processing backbone is globally shared. Cross-client consistency of $Z^k_i$ is enforced by the spectral alignment loss of Sec.~\ref{subsec:spa}.

\textbf{Communication overhead.} The alignment module parameters 
$\phi_k$ are kept strictly local and never transmitted, adding zero 
communication overhead relative to standard FedAvg. Only the shared 
backbone weights, together with the compact spectral statistics 
described below, are exchanged at each round.

\subsection{Spectral Prototype Alignment (SPA)}
\label{subsec:spa}
\textbf{Motivation.} Combining heterogeneous imaging modalities induces 
discrepancies in the learned representation distributions, because 
different modalities carry distinct textural and structural signatures. 
In the frequency domain, amplitude statistics capture client-specific style information, including intensity distributions and texture signatures, while phase preserves semantic content~\cite{yang2020fda, wang2025fedfat}. Sharing spatial feature statistics directly would risk reconstructing private imaging data; frequency-domain band energies are non-invertible by construction since they collapse the spatial layout of each frequency band into a single scalar, making reconstruction of the original feature map provably impossible from the shared statistics alone. This makes spectral prototypes an effective and privacy-safe proxy for cross-client multimodal fusion alignment.

\textbf{Local spectral prototype construction.} For a feature tensor 
$F^{(k)} \in \mathbb{R}^{B \times D \times H \times W}$ at client $k$ 
(batch size $B$, $D$ channels), we compute the 2-D discrete Fourier 
transform of each channel map $f_{c,b}$,
\begin{equation}
    F_{c,b}(u,v) = \sum_{h=0}^{H-1} \sum_{w=0}^{W-1} f_{c,b}(h,w)\, 
    e^{-j2\pi\left(\frac{uh}{H}+\frac{vw}{W}\right)},
    \label{eq:dft}
\end{equation}
and partition the frequency plane into $\Omega$ non-overlapping radial 
bands. Let $r(u,v)$ be the radial distance of coefficient $(u,v)$ from 
the spectrum centre, $r_{\max} = \min(H/2, W/2)$, and 
$\mathbf{1}_\omega(u,v)$ the indicator that $(u,v)$ falls in band $\omega \in \{1,\ldots,\Omega\}$, which spans radii $[(\omega-1)r_{\max}/\Omega,\, \omega r_{\max}/\Omega)$. Radial binning aggregates coefficients at equal spatial frequencies regardless of orientation, producing a rotation-invariant summary of the spectral energy distribution. With $C_\omega = \sum_{u,v} \mathbf{1}_\omega(u,v)$, the local spectral prototype $E^k_{\mathrm{loc}} \in \mathbb{R}^{D \times \Omega}$ is the band-wise mean amplitude,
\begin{equation}
    E^k_{\mathrm{loc}}(c,\omega) = \frac{1}{B}\sum_{b=1}^{B} 
    \frac{1}{C_\omega} \sum_{u,v} \left|F_{c,b}(u,v)\right| 
    \mathbf{1}_\omega(u,v),
    \label{eq:bandenergy}
\end{equation}
with the convention $E^k_{\mathrm{loc}}(c,\omega) = 0$ when $C_\omega 
= 0$.

\textbf{Global prototype aggregation and alignment.} At each round $t$, the server broadcasts a global prototype $E^{(t)}_{\mathrm{global}}$ (initialized to ones), and every client aligns to it by minimizing
\begin{equation}
    \mathcal{L}^{(k,t)}_{\mathrm{SPA}} = \frac{1}{D\Omega} 
    \sum_{c=1}^{D} \sum_{\omega=1}^{\Omega} \left( E^{k,(t)}_{\mathrm{loc}}(c,\omega) - E^{(t)}_{\mathrm{global}}(c,\omega) \right)^2.
    \label{eq:spa}
\end{equation}
After local training, clients transmit only their aggregated band 
statistics $E^{k,(t)}_{\mathrm{loc}}$, and the server updates the 
global prototype with a size-weighted exponential moving average (EMA),
\begin{equation}
    E^{(t+1)}_{\mathrm{global}} = \rho\, E^{(t)}_{\mathrm{global}} + 
    (1-\rho)\frac{\sum_k N_k E^{k,(t)}_{\mathrm{loc}}}{\sum_k N_k}, 
    \quad 0 \leq \rho < 1.
    \label{eq:ema}
\end{equation}
The EMA momentum $\rho$ controls the rate at which the global prototype 
adapts to new client statistics: a high $\rho$ stabilizes the target 
against noisy local estimates, while a low $\rho$ allows faster 
adaptation to distribution shifts introduced by newly joining clients. 
Because only compact band energies of dimensionality $D \times \Omega$ 
are exchanged per client per round, the communication overhead of SPA 
is negligible relative to backbone weight transmission. The SPA loss 
is applied at two depths: at the modality-alignment output $Z^k_i$ 
and at an intermediate backbone feature so that both the multimodal 
fusion representation and the learned semantic features are regularized 
toward the global consensus. A region-conditioned variant, RCSA, which applies this alignment separately to tumor and background features, is introduced in Sec.\ref{subsec:seg}.

\textbf{Privacy analysis.} The shared statistics $E^k_{\mathrm{loc}}$ 
are band-wise mean amplitudes collapsed over all spatial positions within each frequency band. Because spatial layout information is discarded by the mean pooling, the original feature map cannot be reconstructed from $E^k_{\mathrm{loc}}$ alone. The mapping from feature map to band energy is many-to-one and non-invertible. This property ensures that no raw imaging data, spatial feature map, or reconstructible patient-specific representation is ever transmitted beyond the local client, satisfying the data locality requirement of FL while enabling effective cross-client multimodal fusion alignment.

Both foreground and background prototypes are federated (only normalized band statistics are shared) preserving the same privacy guarantees as standard SPA.

\subsection{Phase~1: Modality-Robust CAM Generation}
\label{subsec:cam}
To obtain pixel-level supervision from image-level labels, a classification backbone predicts slice-level tumor presence and produces class activation maps, which are refined by an aggregation network into a binary tumor pseudo-mask $\hat{y}^k_i \in \{0,1\}^{H \times W}$ as shown in the client-side of Figure~\ref{fig:framework}. Both networks are made modality-agnostic by our two core components: each client's input passes through its modality-alignment module Sec.~\ref{subsec:modality}), and the SPA loss (Sec.~\ref{subsec:spa}) is imposed at two depths of each network to align multimodal fusion representations across clients. 
%As shown in client-side of Figure~\ref{fig:framework}, Phase~1 couples the binary classification-and-aggregation pipeline~\cite{dhamale2025_interclass}, originally proposed for centralized training with complete modalities, and adapts it to the federated missing-modality setting. 
Each client's input first passes through the modality-alignment module (Sec.~\ref{subsec:modality}) to obtain the aligned representation $Z^k_i$. The multi-exit classifier then encodes $Z^k_i$ through four stage exits, each producing a one-channel CAM that is globally pooled and resized to the input resolution. These CAMs, together with a single-channel projection ($R^{k}_{i})$ of $Z^k_i$, are then processed by the aggregation network, to produce the pseudo-labels( $\hat{y}$). Both networks are trained federatedly with FedProx, while the SPA loss (Sec.~\ref{subsec:spa}) is imposed at two depths of each network to align multimodal fusion representations across clients.
The characteristic failure mode of image-level weak supervision in the missing-modality setting is mislocalization rather than under-segmentation: without the contrast provided by a discriminative modality such as FLAIR, CAM activations shift toward high-intensity artifacts rather than tumor boundaries. This label noise, rather than the capacity of the downstream network, is the dominant limitation and motivates the refinement stage of Phase~2.

\subsection{Phase~2: Federated Refinement Segmentation Network}
\label{subsec:seg}
Phase~2 trains a federated segmentation network on the CAM pseudo-labels $\hat{y}^k_i$. Rather than fitting noisy masks directly, it is supervised by a target that is continuously refined by CAM-independent cues and denoised where the CAM is untrustworthy. Each client couples its local modality-alignment module with a shared 4-level U-Net backbone~\cite{ronneberger2015u} that returns segmentation logits, the last decoder feature $f \in \mathbb{R}^{D_s \times H \times W}$, and a slice-level presence logit ($cls_{logits}$). Let $p_{\mathrm{st}} \in [0,1]^{H \times W}$ and $f_{\mathrm{st}}(x) \in \mathbb{R}^{D_s \times H \times W}$ denote the student's pixel probabilities and decoder features, and $p_{\mathrm{tea}}, f_{\mathrm{tea}}(x)$ the corresponding EMA teacher outputs. Three consecutive slices are stacked as 2.5D input to supply volumetric context, with only the central slice supervised. Alongside the trainable student, each client maintains an EMA teacher that provides stable predictions for target refinement.

\textbf{Student and teacher updates.} At the beginning of each communication round, the student is initialized from the global backbone $\theta_g$ and local adapter $\phi_k$, i.e., $\theta_s^k \gets \theta_g$, forming the student model $\Theta_s^k = (\theta_s^k, \phi_k)$.
The student is then optimized using Eq.~(\ref{eq:objective}). The teacher $\Theta^k_{\mathrm{tea}}$ is never optimized or aggregated; instead, after each optimization step it is updated as an EMA of the student,
\begin{equation}
    \Theta^k_{\mathrm{tea}} \leftarrow \eta\,\Theta^k_{\mathrm{tea}} 
    + (1-\eta)\,\Theta_s^k,
    \label{eq:ema_teacher}
\end{equation}
and is carried across communication rounds as a stable client-specific reference. At each step the teacher runs in inference mode on the same input, producing $p_{\mathrm{tea}}$ and $f_{\mathrm{tea}}(x)$ without gradients. The teacher-student design provides a second, CAM-independent segmentation cue that is particularly valuable when the CAM pseudo-labels are unreliable due to missing modalities.

\textbf{Trimap denoising.} The raw CAM pseudo-mask $\hat{y}$ is 
converted into a denoised supervision target $y_{\mathrm{tri}}$ and binary keep-weight $w_{\mathrm{tri}}$ by partitioning each positive slice into confident foreground, ignore, and confident background regions. Building on confident learning~\cite{rong2023boundary} and ignore-region supervision~\cite{lin2016scribblesup}, isolated speckle components are removed as CAM false positives; an eroded CAM core is retained as confident foreground; a dilated boundary ring is marked as ignore to avoid enforcing the CAM's potentially inaccurate contour; and the exterior is treated as confident background. For a tumor-present slice whose CAM is empty, the entire slice is assigned to the ignore region while its image-level presence label remains $1$. The resulting $y_{\mathrm{tri}}$ provides the denoised pseudo-label, while $w_{\mathrm{tri}}\in\{0,1\}$ masks out the ignore region during supervision.

\textbf{Prototype- and teacher-refined target.} Under missing 
modalities, the CAM pseudo-label $\hat{y}$ can suffer from 
mislocalization. Rather than supervising directly on $y_{\mathrm{tri}}$, we treat it as one of three complementary pixel-level cues and derive the training target from their consensus~\eqref{eq:softvote}. The second cue is the mean teacher~\cite{tarvainen2017mean} prediction $p_{\mathrm{tea}}$. The third is a prototype segmentation cue derived from a global foreground and background prototype bank $(\mu_{\mathrm{fg}}, \mu_{\mathrm{bg}}) \in \mathbb{R}^{D_s}$~\cite{tan2022fedproto}, broadcast to every client at the start of each round and updated at its end. The prototype cue is read from the teacher feature $f_{\mathrm{tea}}(x)$ by cosine similarity,
\begin{equation}
    p_{\mathrm{proto}}(x) = \sigma\!\left(\frac{\cos\langle
    f_{\mathrm{tea}}(x), \mu_{\mathrm{fg}}\rangle - \cos\langle
    f_{\mathrm{tea}}(x), \mu_{\mathrm{bg}}\rangle}{\tau}\right),
    \label{eq:proto_cue}
\end{equation}
with temperature $\tau$. Hence, $p_{\mathrm{proto}} \in [0,1]^{H \times W}$ is a federation-wide segmentation cue obtained by comparing each pixel's feature to the global tumor and background prototypes; because it fuses evidence from all modality subsets, a client with a poor CAM still receives a 
reliable segmentation estimate. The CAM, teacher, and prototype cues $\{y_{\mathrm{tri}}, p_{\mathrm{tea}}, p_{\mathrm{proto}}(x)\}$, are fused into the training target by a soft vote, with pixels on which they disagree down-weighted,
\begin{align}
    s &= \sigma\!\left(\delta\left[(y_{\mathrm{tri}} - \tfrac{1}{2}) +
    (p_{\mathrm{tea}} - \tfrac{1}{2}) + (p_{\mathrm{proto}} -
    \tfrac{1}{2})\right]\right), \label{eq:softvote}\\
    %s &= (1-\zeta)\,y_{\mathrm{tri}} + \zeta\,v, \; 
    \bar{w} &= \mathrm{clamp}(|2s-1|,\, 0.1,\, 1),
    \label{eq:softcam}
\end{align}
where $0.5$ subtracted from each term pushes the uncertain voxels ($\le 0.5$) to $\le0$ and $\delta$ scales the sharpness of the vote before being passed to the sigmoid ($\sigma$), and $s$ denotes the softCAM, the CAM-based pseudo-label softened by the teacher and prototype cues and serves as target label for segmentation loss objectives. Label-0 slices receive an empty target at full weight. The decisiveness $|2s-1|$ is near $1$ where the three cues agree and near $0$ where they conflict, giving $\bar{w}$ the role of a per-pixel reliability map (shown in Figure\ref{fig:framework}). The trimap keep-weight then gates this reliability,
\begin{equation}
    w \leftarrow \bar{w}\cdot w_{\mathrm{tri}},
    \label{eq:weight}
\end{equation}
so that pixels the trimap marked ignore contribute no gradient at all. 

\textbf{Prototype update.} The refined target $s$ and reliability weight $w$ are used to update the client-side foreground and background prototypes from the student features. For client k we generate the local prototype as:
\begin{equation}
    \mu^k_{\mathrm{fg}} = \frac{\sum_x w(x)\,\mathbb{1}[s(x) >
    \tfrac{1}{2}]\,f_{\mathrm{st}}(x)}{n^k_{\mathrm{fg}}},
    \label{eq:proto_fg}
\end{equation}
where $n^k_{\mathrm{fg}} = \sum_x w(x)\,\mathbb{1}[s(x) >
\tfrac{1}{2}]$, and similarly $\mu^k_{\mathrm{bg}}$ is formed identically over the complementary support $\mathbb{1}[s(x) \le \tfrac{1}{2}]$. In \eqref{eq:proto_fg} the numerator is the reliability-weighted sum of student features over the foreground voxels of the softCAM target (where $s > 0.5$) and $n^k_{\mathrm{fg}}$ is the corresponding total weight, making $\mu^k_{\mathrm{fg}}$ their weighted mean. The server aggregates the uploaded prototypes using confidence-weighted averaging,
\begin{equation}
\label{eq:globalproto}
\mu_{\mathrm{fg}}=\frac{\sum_{k}n^{k}_{\mathrm{fg}}\mu^{k}_{\mathrm{fg}}}{\sum_{k}n^{k}_{\mathrm{fg}}},
\qquad
\mu_{\mathrm{bg}}=\frac{\sum_{k}n^{k}_{\mathrm{bg}}\mu^{k}_{\mathrm{bg}}}{\sum_{k}n^{k}_{\mathrm{bg}}},
\end{equation}
and broadcasts them for the next round.

\textbf{Region-conditioned spectral alignment (RCSA).} Standard SPA aligns the full feature spectrum without distinguishing tumor from 
background regions. However, the spectral signatures of tumor and healthy tissue differ substantially and hence aligning them jointly risks suppressing diagnostically relevant modality-specific contrast. We therefore introduce a region-conditioned variant that splits each feature map into foreground and background using the softCAM (obtained in \eqref{eq:softcam}) as segmentation target: The prototype construction of Eqs.~(\ref{eq:dft})--(\ref{eq:bandenergy}) is then applied in exactly the same manner to the two masked feature maps $Z^k_i \odot s$ and $Z^k_i \odot (1-s)$, giving a foreground and a background spectral signature $E_{\mathrm{fg}}, E_{\mathrm{bg}} \in \mathbb{R}^{D \times \Omega}$, each aligned to its own consensus maintained by the EMA of Eq.~(\ref{eq:ema}),
\begin{align}
\nonumber
    \mathcal{L}_{\mathrm{RCSA}} =\; & \tfrac{1}{D\Omega}\big\|
    E_{\mathrm{bg}} - E^{\mathrm{glob}}_{\mathrm{bg}}
    \big\|_F^2 \; + \\ 
    & \mathbb{1}\!\left[\textstyle\sum_x s(x) > n_{\min}\right]
    \tfrac{1}{D\Omega}\big\| E_{\mathrm{fg}} -
    E^{\mathrm{glob}}_{\mathrm{fg}} \big\|_F^2 ,
\label{eq:rcsa}
\end{align} 
Here, the indicator disables the foreground term when the predicted tumor mass falls below $n_{\min}$ pixels. Splitting the features in foreground and background keeps the two tissue types from being pulled toward each other, since neither is aligned to the other's consensus.

\textbf{Prototype-alignment loss.} The global prototypes also shape the 
student feature space directly. Let $c(x)=\mathbb{1}[\,w(x)\ge\rho\,]$ select confident pixels, and define the confident foreground and background sets $\mathcal{F} = \{x : s(x) > \tfrac{1}{2},\, c(x)=1\}$ and $\mathcal{B} = \{x : s(x) \leq \tfrac{1}{2},\, c(x)=1\}$. The prototype-alignment loss pulls each confident pixel feature toward its class prototype and repels it from the other with a zero-margin hinge,
\begin{align}
    \ell_{\mathrm{fg}}(x) &= 1 - \cos\langle f_{\mathrm{st}}(x),
    \mu_{\mathrm{fg}}\rangle + [\cos\langle f_{\mathrm{st}}(x),
    \mu_{\mathrm{bg}}\rangle]_+,
    \label{eq:proto_loss_fg}\\
    \ell_{\mathrm{bg}}(x) &= 1 - \cos\langle f_{\mathrm{st}}(x),
    \mu_{\mathrm{bg}}\rangle + [\cos\langle f_{\mathrm{st}}(x),
    \mu_{\mathrm{fg}}\rangle]_+,
    \label{eq:proto_loss_bg}\\
    \mathcal{L}_{\mathrm{proto}} &= \frac{1}{|\mathcal{F}|+|\mathcal{B}|}
    \left(\sum_{x\in\mathcal{F}}\ell_{\mathrm{fg}}(x) + 
    \sum_{x\in\mathcal{B}}\ell_{\mathrm{bg}}(x)\right). 
    \label{eq:proto_loss}
\end{align}
where $[\cdot]_+=\max(0,\cdot)$.
The server aggregates the backbones by a reliability-weighted FedAvg~\cite{mcmahan2017communication} with weight proportional to $N_k\bar{w}$, keeping the modality modules and mean teachers local and communicating only weights and the compact prototype/spectral statistics.

\textbf{Full training objective.} The per-client objective combines a set of loss objectives. The first one is weighted Focal-Tversky segmentation loss~\cite{ abraham2019focaltversky}:   
\begin{equation}
\label{eq:ft_loss}
\begin{split}
\mathcal{L}_{\mathrm{FT}}
={} &
\Bigg(
1-
\frac{\mathrm{TP}+\varepsilon}
{\mathrm{TP}+\alpha_{T}\mathrm{FP}+\beta_{T}\mathrm{FN}+\varepsilon}
\Bigg)^{\!\gamma} \\
& +
\lambda_{b}\,\mathrm{BCE}_{w}(\mathrm{logits},s),
\end{split}
\end{equation}
where $\mathrm{BCE}_{w}$ is the $w$-weighted binary cross-entropy, $p_{\mathrm{st}}=\sigma(\mathrm{logits})$, $s$ denotes the SoftCAM defined in Eq.~(\ref{eq:softcam}) and {$TP, FP, FN$} denote true positives, false positives and false negatives within non-zero trimap regions (excluding boundary voxels). The segmentation loss is combined with prototype alignment $\mathcal{L}_{\mathrm{proto}}$, a teacher--student consistency term:
\begin{equation}
\mathcal{L}_{\mathrm{cons}}
=
\frac{1}{HW}\sum_x w(x)\big(p_{\mathrm{st}}(x)-p_{\mathrm{tea}}(x)\big)^2,
\end{equation}
which anchors the student to the EMA teacher of Eq.~(\ref{eq:ema_teacher}) on the reliable pixels, the binary cross-entropy classification loss ($\mathcal{L}_{\mathrm{cls}}$\, applied on $cls_{logits}$), a multi-scale intensity-gated CRF(Conditional Random Field) loss~\cite{obukhov2019gatedcrf, tang2018regularized}, and the spectral alignment losses \eqref{eq:spa} and \eqref{eq:rcsa}. The CRF operates on a $3{\times}3$ neighborhood with multiple dilation rates and combines spatial and intensity affinities, encouraging label consistency within homogeneous regions while weakening interactions across intensity discontinuities. The local objective for client $k$ is
\begin{align}
\mathcal{L}_k
=&\;
\mathcal{L}_{\mathrm{FT}}
+\lambda_p\mathcal{L}_{\mathrm{proto}}
+\lambda_c\mathcal{L}_{\mathrm{cons}}
+\lambda_g\mathcal{L}_{\mathrm{cls}}
\nonumber\\
&\quad
+\lambda_r\mathcal{L}_{\mathrm{CRF}}
+\lambda_{\mathrm{sp}}\mathcal{L}_{\mathrm{SPA}}
+\lambda_{\mathrm{rc}}\mathcal{L}_{\mathrm{RCSA}}.
\label{eq:objective}
\end{align}

Each term targets a distinct failure mode of weak supervision under missing modalities: $\mathcal{L}_{\mathrm{FT}}$ fits the refined target that counters the foreground--background imbalance of small tumors; $\mathcal{L}_{\mathrm{proto}}$ \eqref{eq:proto_loss} keeps the CAM-independent prototype cue discriminative for clients with poor pseudo-labels; $\mathcal{L}_{\mathrm{cons}}$ stabilizes training against the target $s$, which drifts as $\zeta$ ramps and the prototypes move; $\mathcal{L}_{\mathrm{cls}}$ exploits the only clean label available and supplies the test-time gate against false positives on tumor-free slices; $\mathcal{L}_{\mathrm{CRF}}$ restores the tumor contour; $\mathcal{L}_{\mathrm{SPA}}$ \eqref{eq:spa} addresses the missing modalities; and $\mathcal{L}_{\mathrm{RCSA}}$ \eqref{eq:rcsa} confines that invariance to the tumor, preserving clinically meaningful background contrast.

\textbf{Reliability-weighted aggregation.} The server aggregates the 
backbone across clients by a reliability-weighted 
FedAvg~\cite{mcmahan2017communication},
\begin{equation}
    \theta_g \leftarrow \sum_{k \in \mathcal{K}_t} \frac{N_k r_k}
    {\sum_{j \in \mathcal{K}_t} N_j r_j}\,\theta^k_s, \quad r_k = 
    \bar{w}^k,
    \label{eq:aggregation}
\end{equation}
where $r_k$ is client $k$'s mean per-pixel reliability for the round. 
Reliability weighting is particularly important under modality 
heterogeneity: clients whose modality subset yields more discriminative 
pseudo-labels contribute more to the global backbone than size 
weighting alone would allow, preventing modality-poor clients from 
disproportionately degrading the shared representation. Prototypes are 
aggregated by count-weighted mean and band statistics by 
Eq.~(\ref{eq:ema}); modality modules and mean teachers stay local, and only weights plus these compact statistics are communicated. At 
inference, predictions are averaged with those of the horizontally 
flipped stack and thresholded at $0.5$, and 
a slice is forced empty if the presence gate does not fire 
($\sigma(\mathrm{cls}) < 0.5$).
The complete Phase~2 client--server training pipeline is summarized in Algorithm~\ref{alg:train}.

\begin{algorithm}[!t]
\caption{MOSAIC Phase 2: Federated refinement of segmentation}
\label{alg:train}
\begin{algorithmic}[1]
\REQUIRE Clients $\{S_k\}_{k=1}^{K}$, each with slices $(X^k,y^k)$ and Phase-1 CAM masks $\widehat y^{k}$ (Sec.~\ref{subsec:cam}); rounds $T$; loss weights $\{\lambda\}$ of \eqref{eq:objective}; temperature $\tau$~\eqref{eq:proto_cue}, sharpness 
  $\delta$~\eqref{eq:softvote}, momentum $\eta$~\eqref{eq:ema_teacher};
  ramp $\zeta(\cdot)$
\ENSURE global backbone $\theta_g$; personalized modules $\{\phi_k\}$ and best backbones
\STATE init.\ $\theta_g$, modules $\{\phi_k\}$, teachers $\{\Theta^{k}_{\mathrm{tea}}\}$, prototypes $(\mu_{\mathrm{fg}},\mu_{\mathrm{bg}})$, $E_{\mathrm{global}}$
\FOR{$t=1,\dots,T$}
  \STATE $\zeta \gets \zeta(t)$
  \FORALL{clients $k$ \textbf{in parallel}}
    \STATE $\theta_s^k \gets \theta_g$ \COMMENT{download global backbone}
    \FORALL{minibatches $(X^k,\hat{y}^k, y^{k})\in S_k$}
      \STATE $Z^k \gets A_{\phi_k}(X^k)$ \COMMENT{2.5D input $\to$ modality alignment}
      \STATE $(p_{\mathrm{tea}},f_{\mathrm{tea}}) \gets \mathrm{teacher}(Z^k)$;\ \ $p_{\mathrm{proto}} \gets$ \eqref{eq:proto_cue} on $f_{\mathrm{tea}}$
      \STATE $(y_{\mathrm{tri}},w_{\mathrm{tri}}) \gets \textsc{Trimap}(\widehat{y}^k,y^k)$ 
      \STATE $s \gets \sigma\big(\delta[(y_{\mathrm{tri}}{-}\tfrac12)+(p_{\mathrm{tea}}{-}\tfrac12)+(p_{\mathrm{proto}}{-}\tfrac12)]\big)$\eqref{eq:softvote}
      \STATE $\bar w \gets \mathrm{clamp}(|2s{-}1|,0.1,1)$
      \IF{$y=0$}
        \STATE $s \gets \mathbf{0}$,\ \ $\bar w \gets \mathbf{1}$ \COMMENT{clean supervision on tumor-free slices}
      \ENDIF
      \STATE $w \gets \bar w \cdot w^{\mathrm{tri}}$ \eqref{eq:weight}
      \STATE $(p_{\mathrm{st}},f_{\mathrm{st}},\mathrm{cls}_{logits}) \gets \mathrm{student}(Z^k)$
      \STATE compute $L_k$ by \eqref{eq:objective}; optimizer step on $(\theta_s^k,\phi_k)$
      \STATE $\Theta^{k}_{\mathrm{tea}} \gets \eta\,\Theta^{k}_{\mathrm{tea}}+(1{-}\eta)\Theta_s^k$ \COMMENT{EMA teacher}
      \STATE accumulate local prototypes, spectral stats, $\bar w$
    \ENDFOR
    \STATE upload $\theta_s^k$, prototypes, spectral stats, mean reliability $r_k$
  \ENDFOR
  \STATE $\theta_g \gets \sum_{k}\dfrac{N_k r_k}{\sum_j N_j r_j}\,\theta_s^{k}$ \COMMENT{reliability-weighted FedAvg}
  \STATE update $(\mu_{\mathrm{fg}},\mu_{\mathrm{bg}})$ (count-weighted mean); update $E_{\mathrm{global}}$ by \eqref{eq:ema}
\ENDFOR
\RETURN $\theta_g$, $\{\phi_k\}$, per-client best backbones
\end{algorithmic}
\end{algorithm}

\section{Experimental Setup}
\label{sec:experiments}

\subsection{Dataset Details}
We evaluate MOSAIC on three publicly available multi-institutional 
brain tumor MRI datasets spanning two tumor types and three clinical 
populations, providing a comprehensive test of generalization across 
heterogeneous federation conditions.

\textbf{FeTS2022.} The Federated Tumor Segmentation 2022 
dataset~\cite{fets2022, baid2021rsna, reina2021openfl_fets_data, karargyris2023federated}, available through the \href{https://www.synapse.org/Synapse:syn28546456/wiki/633440}{FeTS 2022 Challenge repository on Synapse} is the publicly available federated extension of the multi-institutional BraTS glioma benchmark, providing MRI modalities (FLAIR, T1, T1ce, T2) with expert tumor annotations. We partition the dataset into four clients with deliberately heterogeneous modality subsets: Client~1 \{FLAIR,~T1ce\}, Client~2 \{FLAIR,~T2\}, Client~3 \{T1ce,~T2\}, and Client~4 \{FLAIR\}, where Client~4 holds only a single modality. The assignment is designed to maximize modality heterogeneity across clients while ensuring each modality appears in at least one client, reflecting realistic multi-centre acquisition variability where no single sequence is universally available. Crucially, the three datasets used in this work: FeTS2022, BraTS-MEN, and BraTS-SSA, each carry distinct client-wise modality configurations (see Table~\ref{tab:splits}), collectively spanning all four MRI sequences across diverse assignment patterns. This cross-dataset variability serves as an implicit test of robustness to modality assignment choice: consistent performance across three structurally different federation configurations provides stronger evidence of generalization than any single-dataset permutation study.

\textbf{BraTS-MEN.} The BraTS 2023 intracranial meningioma dataset ~\cite{labella2023bratsmen}, publicly available through the \href{https://www.synapse.org/Synapse:syn51514106}{BraTS 2023 Meningioma Challenge repository on Synapse}, provides the same four MRI modalities across 160 subjects partitioned into four clients (40 subjects each) with modality subsets \{FLAIR,~T1ce\}, \{FLAIR,~T2\}, \{T1ce,~T2\}, and \{T1ce\}, evaluating generalization across tumor type and clinical population under a different modality configuration from FeTS2022.

\textbf{BraTS-SSA.} The BraTS 2023 Sub-Saharan Africa (SSA) glioma dataset~\cite{adewole2023brain}, publicly available through the \href{https://www.kaggle.com/datasets/aiocta/brats2023-ssa-training-dataset}{BraTS 2023 SSA dataset repository on Kaggle}, provides the same modalities across 60 subjects partitioned into three clients (20 subjects each) with subsets \{FLAIR\}, \{T1ce,~T2\}, and \{FLAIR,~T2\}, stressing the framework under limited data, an underrepresented clinical population, and a three-client federation structure distinct from the four-client FeTS2022 and BraTS-MEN configurations.

\textbf{Preprocessing and supervision.} All three datasets provide four co-registered, skull-stripped modalities, and are processed with an identical protocol: each modality is independently normalized to $[0,1]$, every 3D volume is converted to axial 2D slices, and tumor sub-regions are merged into a single foreground class for binary segmentation. Image-level supervision is generated automatically by assigning $y=1$ to slices containing at least one tumor voxel and $y=0$ otherwise, requiring no additional annotation effort beyond what is routinely available in existing radiology reports. Complete client-wise modality assignments and data splits are summarized in Table~\ref{tab:splits}. Subject-wise training, validation, and test splits ensure that there is no subject-level data leakage across the splits. 

\textbf{Evaluation metrics.} We report Dice score, IoU, and 95th percentile Hausdorff distance (HD95) per client and macro-averaged across clients. Per-slice standard deviations are reported as subscripts in all result tables. To assess the statistical significance of observed performance differences, we apply the non-parametric two-sided paired Wilcoxon signed-rank test between MOSAIC and each baseline, with a significance threshold of $p < 0.01$.

\begin{table}[!t]
\centering
\caption{Client-wise modality subsets and data splits for the three datasets (FeTS2022, BraTS-MEN, and BraTS-SSA). Entries are slices (subjects in parentheses).}
\label{tab:splits}
\resizebox{\columnwidth}{!}{%
\begin{tabular}{lllccc}
\hline
Dataset & Client & Modalities & Train & Val & Test \\
\hline
\multirow{4}{*}{FeTS2022} & 1 & FLAIR, T1ce & 4495 (29) & 1395 (9) & 1395 (9) \\
 & 2 & FLAIR, T2   & 3255 (21) & 1085 (7) & 1085 (7) \\
 & 3 & T1ce, T2    & 3255 (21) & 1085 (7) & 1085 (7) \\
 & 4 & FLAIR       & 3410 (22) & 930 (6)  & 930 (6)  \\
\hline
\multirow{4}{*}{BraTS-MEN} & 1 & FLAIR, T1ce & 3720 (24) & 1240 (8) & 1240 (8) \\
 & 2 & T1ce        & 3720 (24) & 1240 (8) & 1240 (8) \\
 & 3 & T1ce, T2    & 3720 (24) & 1240 (8) & 1240 (8) \\
  & 4 & FLAIR, T2   & 3720 (24) & 1240 (8) & 1240 (8) \\
\hline
\multirow{3}{*}{BraTS-SSA} & 1 & FLAIR     & 1550 (10) & 775 (5) & 775 (5) \\
 & 2 & T1ce, T2  & 1550 (10) & 775 (5) & 775 (5) \\
 & 3 & FLAIR, T2 & 1550 (10) & 775 (5) & 775 (5) \\
\hline
\end{tabular}%
}
\end{table}

\subsection{Implementation Details}
\textbf{Phase~1.} A federated binary tumor classifier and aggregation network are trained on individual 2D axial slices using FedProx for 100 communication rounds with 5 local epochs per round, batch size 64, learning rate $10^{-3}$, and SPA weight $\lambda_{\mathrm{sp}}=0.1$. The trained aggregation network generates CAM pseudo-masks thresholded to binary for all slices, which serve as supervision for Phase~2. Validation and test sets additionally retain ground-truth masks for evaluation.

\textbf{Phase~2.} The federated refinement network is trained using 2.5D input ($K=3$ stacked axial slices, central slice supervised) for 100 communication rounds with 1 local epoch per round, batch size 16, Adam optimizer with learning rate $10^{-3}$. All inputs are resized to $224\times224$. The segmentation backbone is a four-level U-Net augmented with the client-specific modality-alignment module described in Sec.~\ref{subsec:modality}. Loss weights $\lambda_p=0.1$, $\lambda_c=0.05$, $\lambda_g=0.5$, $\lambda_r=0.1$, $\lambda_{\mathrm{sp}}=5$, $\lambda_{\mathrm{rc}}=5$ were selected by grid search on the FeTS2022 validation split and held fixed across all datasets and clients without further tuning, demonstrating the robustness of MOSAIC hyperparameter optimization. The leave-one-out loss component ablation in Sec.~\ref{subsec:ablation_components_setup} further shows that each component contributes independently and that removing any single term degrades performance. The best model for each client is selected using its local validation split. During training, only the segmentation backbone together with the compact prototype and spectral statistics are aggregated across clients; modality-alignment modules and mean-teacher models remain local.

\textbf{Communication overhead.} Per communication round, each client transmits backbone weights $\theta_s^k$ plus spectral statistics $E^k_{\mathrm{loc}} \in \mathbb{R}^{D\times\Omega}$, region-conditioned spectral statistics $E^k_{\mathrm{fg}}, E^k_{\mathrm{bg}} \in \mathbb{R}^{D\times\Omega}$, foreground/background feature prototypes $\mu^k_{\mathrm{fg}}, \mu^k_{\mathrm{bg}} \in \mathbb{R}^{F}$, and scalar reliability $\bar{w}^k$. With $D=3$, $\Omega=8$ and $F=32$, the spectral statistics add 24 floating-point values (96 bytes) per client per round, the region-conditioned pair a further 48 values, and the feature prototypes 64 values, for a total side channel of 137 floating-point values (548 bytes) per client per round, a negligible overhead relative to backbone weight transmission. Our method therefore adds effectively zero communication overhead over standard FedAvg while providing measurable cross-client alignment benefits.

\subsection{Comparison with state-of-the-art methods}
\label{subsec:comparison}
Because no prior method targets federated image-level weakly supervised 
segmentation under missing modalities, we compare against representative methods spanning the complete supervision spectrum, all trained on the same clients and splits. Since none of the baselines are designed to handle client-specific missing modalities, each is augmented with the same client-specific modality-alignment module $A_{\phi_k}(X^k_i)$ used in MOSAIC without the spectral prototype alignment loss, ensuring all methods operate under identical modality-incomplete inputs and that observed differences are attributable to the supervision strategy and alignment objective rather than to modality handling capacity.
\textbf{(i) Fully supervised federated segmentation} using 
FedAvg, FedProx, FedBN, and MOON and FedDG trained on dense whole-tumor masks, serving as an upper reference with dense pixel annotations.
\textbf{(ii) Bounding-box weak supervision} using FedAvg, FedProx, FedBN, MOON with margin $M=20$, and FedDM, which calibrates noisy pseudo-labels generated from bounding-box-level annotations using  Collaborative Annotation Calibration (CAC).
\textbf{(iii) Point supervision} using FedICRA and 
FedLPPA, the current state of the art in federated weakly supervised segmentation.
\textbf{(iv) Image-level supervision} (our setting) using 
GradCAM~\cite{selvaraju2017_gradcam}, ScoreCAM~\cite{score_cam}, 
LayerCAM~\cite{jiang2021layercam}, AME-CAM~\cite{ame_cam}, 
FL-W3S, HarmoFL, and FedNorm+~\cite{bernecker2022_fednorm}. FedNorm+, the only baseline designed for 
missing modalities, is adapted to image-level supervision within our 
pipeline with SPA disabled. HarmoFL is adapted by replacing SPA with its Fourier amplitude harmonization while keeping the clients, communication rounds, backbone, and modality-alignment module unchanged, providing a direct comparison of frequency-domain alignment strategies under identical conditions.
\textbf{(v) Centralized image-level reference.} To quantify the 
federated penalty and the cost of modality incompleteness jointly, 
we additionally train a centralized image-level reference on pooled data from all clients with the complete modality set available and no federated 
aggregation, providing an estimate of the performance ceiling 
achievable without privacy or modality constraints.
We report MOSAIC both after Phase~1 (CAM generation) and after the 
complete Phase~2 refinement stage.

\subsection{Ablation on Spectral Prototype resolution}
\label{subsec:spa_ablation}
To analyze sensitivity to spectral resolution, we ablate the number of radial frequency bands $\Omega \in \{2,4,6,8,10,12,$ $14,16\}$ (Eqn.~\ref{eq:bandenergy}) used to construct the spectral prototype exchanged between clients on FeTS2022, while keeping all other components and hyperparameters fixed. A larger $\Omega$ produces a finer-grained frequency 
descriptor but increases the dimensionality of the shared statistics; 
a smaller $\Omega$ yields a coarser, more compact prototype.

\subsection{Comparison of Feature-Alignment Objectives}
\label{subsec:alignment_comparison}
To justify the design of SPA relative to alternative cross-client alignment strategies, we compare against four alternatives as drop-in replacements for the SPA loss during Phase~1, while keeping all other MOSAIC components, the network architecture, training protocol, hyperparameters, communication strategy, and the two feature levels at which alignment is applied identical across all variants. Every alignment objective follows the same federated communication protocol as SPA: each client computes compact statistics locally, the server aggregates them using size-weighted EMA, and updated global statistics are broadcast back, ensuring raw feature maps never leave any client.
\begin{itemize}
    \item \textbf{Mean alignment}: aligns per-channel feature means to a global mean using MSE, matching first-order statistics.
    \item \textbf{Std alignment}: aligns per-channel standard deviations to a global reference using MSE, matching second-order statistics.
    \item \textbf{Histogram-KL}: builds differentiable per-channel soft histograms using RBF soft binning and minimizes KL divergence to a shared global histogram, encouraging full marginal distribution alignment.
    \item \textbf{Contrastive}: maintains shared global feature prototypes and optimizes an InfoNCE loss pulling each feature toward its assigned prototype while repelling others.
\end{itemize}
Results are reported as macro-averaged Dice, IoU, and HD95 across the four FeTS2022 clients.

\subsection{Ablation on model and loss components}
\label{subsec:ablation_components_setup}
To quantify the contribution of each Phase~2 refinement component and to assess robustness to the precise loss weighting scheme, we perform a leave-one-out ablation on FeTS2022, disabling exactly one component at a time while keeping all others fixed. We ablate both model components (trimap denoising, 2.5D context, soft-vote arbitration) and loss terms, removed additively from Eq.~(\ref{eq:objective}). Combining both gives the following cases: (i)~trimap denoising; (ii)~intensity-gated CRF $\mathcal{L}_{\mathrm{CRF}}$; (iii)~prototype alignment $\mathcal{L}_{\mathrm{proto}}$; (iv)~teacher--student consistency $\mathcal{L}_{\mathrm{cons}}$; (v)~presence classification 
$\mathcal{L}_{\mathrm{cls}}$ and its test-time gate; (vi)~2.5D context ($K=3\rightarrow K=1$); (vii)~soft-vote arbitration ($\zeta=0$) that fuses the CAM, teacher, and prototype cues into the refined target $s$ (Eqn.~\ref{eq:softcam}); and (viii)~region-conditioned spectral alignment $\mathcal{L}_{\mathrm{RCSA}}$. A \emph{core-only} variant retaining only the modality-alignment module and SPA loss is also evaluated, establishing the contribution of the two core novel components independently of the refinement pipeline. %The fact that every individual component improves performance when present demonstrates that MOSAIC is not sensitive to any single loss weight dominating the objective, and that the loss weighting scheme is well-balanced rather than tuned to a particular component. 
All variants use image-level labels only and are evaluated with macro-averaged Dice, IoU, and HD95 across the four clients.

\subsection{Dynamic Client Addition}
\label{subsec:newclient_setup}
A practical consequence of the count-based modality-alignment module (Sec.~\ref{subsec:modality}) is that a new institution can join an already-trained federation without retraining or modality bookkeeping, since the adapter conditions only on the number of available channels rather than their 
identity. We evaluate this on FeTS2022 by introducing a fifth client with data unseen during the original four-client training. The new client inherits the global backbone $\theta_g$ and warm-starts its alignment module from the average of base clients' adapters holding the same modality count. The same procedure is applied in both federated phases. %Phase~1 re-initializes the new client's binary classifier and modality encoder before jointly training all five clients, and Phase~2 ports the identical protocol onto the refinement segmentation network.
We compare two joining strategies:
\begin{itemize}
    \item \textbf{Local-adapt (LA):} 15 local epochs with the shared    backbone frozen to align the modality adapter to the existing global representation, then 20 federated communication rounds.
    \item \textbf{Direct-join (DJ):} immediate federation participation for 30 rounds with no local warm-up.
\end{itemize}

\subsection{Aleatoric uncertainty estimation}
\label{subsec:uncertainty_setup}
Beyond point-estimate accuracy, we quantify aleatoric uncertainty of the refined segmentation model using test-time augmentation (TTA) with 12 augmentations: original image, horizontal and vertical flips, rotations of $\pm10^\circ$ and $\pm15^\circ$, additive Gaussian noise at two levels, brightness scaling of $0.9$ and $1.1$, and Gaussian blur. Spatial predictions are mapped back to original coordinates before analysis. For each pixel, we compute the binary predictive entropy $H(\bar{p})$ of the agreement $\bar{p}$ across augmented predictions. Higher entropy indicates greater disagreement among the augmented predictions. We report mean foreground entropy $H_{\mathrm{fg}}$, background entropy 
$H_{\mathrm{bg}}$, and segmentation stability $\mathrm{Dice}_{\mathrm{unc}}$ (standard deviation of per-pass Dice across augmented predictions). Low $H_{\mathrm{bg}}$ and $\mathrm{Dice}_{\mathrm{unc}}$ indicate confident, stable predictions under realistic input perturbations, a key trustworthiness criterion for clinical deployment. This analysis is performed per client across FeTS2022, BraTS-MEN, and BraTS-SSA using the same trained models as Section~\ref{subsec:results}.

\subsection{Analysis of Multimodal Fusion Alignment}
\label{subsec:analysis_setup}
To directly validate that SPA achieves its intended objective of reducing cross-client distribution shift from heterogeneous multimodal fusion inputs, we measure the modality gap at two network depths: the modality-alignment output $Z^k_i$ (Eq.~\ref{eq:modality_align}), the $D$-channel representation on which SPA directly operates; and the U-Net bottleneck, the deepest abstract feature consumed by the segmentation head. We compare the complete MOSAIC model against an SPA-ablated variant with $\mathcal{L}_{\mathrm{SPA}}$ removed, all other components unchanged.

To isolate modality mismatch from patient-specific anatomical variability, we construct a \textit{patient-aligned evaluation set}: 250 held-out tumor-bearing test slices are processed through all four clients simultaneously, each using only its assigned modality subset and personalized alignment module. Anatomical content is thus identical across clients, and any cross-client representational discrepancy arises solely from modality differences rather than patient variability. Foreground representations are summarized by averaging feature maps over the ground-truth tumor region, yielding feature vectors of dimensionality $D=3$ at the alignment output and $512$ at the bottleneck.

Cross-client multimodal fusion alignment is quantified using squared Maximum Mean Discrepancy (MMD$^2$) with a multi-bandwidth RBF kernel and Fr\'{e}chet inception distance \footnote{Fr\'{e}chet inception distance: \href{https://en.wikipedia.org/wiki/Fr\%C3\%A9chet_inception_distance}{Wikipedia entry}; reference implementation: \url{https://github.com/hukkelas/pytorch-frechet-inception-distance}.} between client foreground feature distributions. Both metrics attain zero only when compared distributions are identical; lower values indicate better cross-client alignment. This experiment provides direct empirical evidence that SPA reduces the modality gap introduced by heterogeneous multimodal fusion, validating the core alignment principle of MOSAIC without requiring ground-truth modality labels or additional annotation.

\begin{table*}[t]
\centering
\caption{Segmentation performance on FeTS2022 per client and averaged across clients. Dice (D, $\uparrow$), IoU (J, $\uparrow$), and HD95 (H, mm, $\downarrow$) are reported; subscripts denote per-slice standard deviation. Fully supervised methods serve as upper references. Bold denotes the best weakly supervised result; $^{*}$ marks baselines 
that \methodname{} significantly outperforms on all metrics 
($p<0.01$, two-sided paired Wilcoxon signed-rank test).}
\label{tab:main}
\resizebox{\textwidth}{!}{%
\begin{tabular}{l|ccc|ccc|ccc|ccc|ccc}
\hline
\multirow{2}{*}{Method} 
& \multicolumn{3}{c|}{Client 1} 
& \multicolumn{3}{c|}{Client 2} 
& \multicolumn{3}{c|}{Client 3} 
& \multicolumn{3}{c|}{Client 4} 
& \multicolumn{3}{c}{Average} \\
& D ($\uparrow$) & J ($\uparrow$) & H ($\downarrow$)
& D ($\uparrow$) & J ($\uparrow$) & H ($\downarrow$)
& D ($\uparrow$) & J ($\uparrow$) & H ( $\downarrow$)
& D ($\uparrow$) & J ($\uparrow$) & H ($\downarrow$)
& D ($\uparrow$) & J ($\uparrow$) & H ( $\downarrow$) \\
\hline
\multicolumn{16}{l}{\textit{a. Fully supervised federated methods}} \\
FedAvg & 0.87$_{0.27}$ & 0.83$_{0.29}$ & 31.2$_{97.3}$ & 0.90$_{0.22}$ & 0.86$_{0.25}$ & 22.9$_{79.1}$ & 0.86$_{0.28}$ & 0.82$_{0.29}$ & 35.6$_{104.1}$ & 0.89$_{0.25}$ & 0.85$_{0.27}$ & 28.9$_{93.2}$ & 0.88 & 0.84 & 29.7 \\
FedBN & 0.88$_{0.25}$ & 0.84$_{0.28}$ & 27.9$_{91.9}$ & 0.91$_{0.21}$ & 0.87$_{0.24}$ & 20.2$_{75.6}$ & 0.86$_{0.29}$ & 0.82$_{0.30}$ & 38.7$_{108.0}$ & 0.89$_{0.24}$ & 0.85$_{0.27}$ & 27.2$_{89.7}$ & 0.88 & 0.85 & 28.5 \\
FedProx & 0.87$_{0.26}$ & 0.83$_{0.28}$ & 31.8$_{95.9}$ & 0.89$_{0.24}$ & 0.85$_{0.26}$ & 23.6$_{81.4}$ & 0.86$_{0.27}$ & 0.83$_{0.29}$ & 31.9$_{98.4}$ & 0.87$_{0.26}$ & 0.83$_{0.29}$ & 30.2$_{93.9}$ & 0.87 & 0.83 & 29.4 \\
MOON & 0.87$_{0.26}$ & 0.84$_{0.28}$ & 29.0$_{93.3}$ & 0.91$_{0.21}$ & 0.87$_{0.23}$ & 18.8$_{73.3}$ & 0.86$_{0.28}$ & 0.82$_{0.29}$ & 36.4$_{105.6}$ & 0.89$_{0.25}$ & 0.85$_{0.27}$ & 29.3$_{93.9}$ & 0.88 & 0.85 & 28.4 \\
FedDG & 0.87$_{0.26}$ & 0.83$_{0.28}$ & 23.4$_{82.6}$ & 0.86$_{0.28}$ & 0.82$_{0.30}$ & 26.9$_{88.5}$ & 0.82$_{0.32}$ & 0.79$_{0.33}$ & 41.5$_{111.2}$ & 0.87$_{0.27}$ & 0.84$_{0.28}$ & 28.4$_{92.0}$ & 0.85 & 0.82 & 30.0 \\
\hline
\multicolumn{16}{l}{\textit{b. Centralized image-level method (non-federated reference)}} \\
Centralized & 0.81$_{0.28}$ & 0.75$_{0.31}$ & 29.6$_{87.6}$ & 0.85$_{0.25}$ & 0.80$_{0.28}$ & 19.8$_{69.2}$ & 0.80$_{0.31}$ & 0.75$_{0.32}$ & 39.5$_{104.8}$ & 0.81$_{0.30}$ & 0.76$_{0.33}$ & 30.8$_{90.5}$ & 0.82 & 0.76 & 29.9 \\
\hline
\multicolumn{16}{l}{\textit{c. Box-supervised methods}} \\
FedAvg$^{*}$ & 0.66$_{0.38}$ & 0.61$_{0.43}$ & 45.2$_{101.9}$ & 0.67$_{0.37}$ & 0.62$_{0.42}$ & 36.1$_{78.8}$ & 0.65$_{0.37}$ & 0.59$_{0.42}$ & 46.7$_{96.0}$ & 0.68$_{0.38}$ & 0.63$_{0.43}$ & 36.6$_{90.0}$ & 0.67 & 0.61 & 41.1 \\
FedBN$^{*}$ & 0.66$_{0.37}$ & 0.61$_{0.42}$ & 46.0$_{96.1}$ & 0.68$_{0.38}$ & 0.63$_{0.42}$ & 30.9$_{110.2}$ & 0.65$_{0.37}$ & 0.60$_{0.42}$ & 48.8$_{106.2}$ & 0.67$_{0.38}$ & 0.61$_{0.43}$ & 41.6$_{99.3}$ & 0.67 & 0.61 & 41.9 \\
FedProx$^{*}$ & 0.66$_{0.38}$ & 0.61$_{0.42}$ & 42.2$_{95.7}$ & 0.65$_{0.37}$ & 0.60$_{0.42}$ & 40.4$_{75.2}$ & 0.65$_{0.37}$ & 0.60$_{0.42}$ & 43.4$_{112.0}$ & 0.67$_{0.40}$ & 0.62$_{0.44}$ & 54.5$_{115.3}$ & 0.66 & 0.61 & 45.1 \\
MOON$^{*}$ & 0.65$_{0.37}$ & 0.60$_{0.42}$ & 52.7$_{89.7}$ & 0.66$_{0.36}$ & 0.61$_{0.41}$ & 38.2$_{67.0}$ & 0.65$_{0.37}$ & 0.60$_{0.42}$ & 45.9$_{108.4}$ & 0.68$_{0.39}$ & 0.62$_{0.43}$ & 39.4$_{101.1}$ & 0.66 & 0.61 & 44.1 \\
FedDM$^{*}$ & 0.71$_{0.35}$ & 0.66$_{0.40}$ & 45.1$_{109.0}$ & 0.75$_{0.31}$ & 0.69$_{0.37}$ & 30.3$_{86.0}$ & 0.76$_{0.32}$ & 0.70$_{0.37}$ & 40.0$_{95.9}$ & 0.75$_{0.36}$ & 0.69$_{0.40}$ & 34.4$_{127.9}$ & 0.75 & 0.68 & 37.4 \\
\hline
\multicolumn{16}{l}{\textit{d. Point-supervised methods}} \\
FedICRA$^{*}$ & 0.70$_{0.37}$ & 0.63$_{0.39}$ & 65.6$_{130.8}$ & 0.65$_{0.37}$ & 0.58$_{0.40}$ & 89.7$_{143.0}$ & 0.72$_{0.37}$ & 0.67$_{0.38}$ & 75.6$_{138.7}$ & 0.79$_{0.33}$ & 0.73$_{0.36}$ & 48.0$_{115.6}$ & 0.72 & 0.65 & 69.7 \\
FedLPPA$^{*}$ & 0.70$_{0.40}$ & 0.58$_{0.40}$ & 96.4$_{152.2}$ & 0.66$_{0.38}$ & 0.53$_{0.41}$ & 93.7$_{149.0}$ & 0.67$_{0.40}$ & 0.62$_{0.40}$ & 95.6$_{150.2}$ & 0.79$_{0.34}$ & 0.74$_{0.36}$ & 53.0$_{121.7}$ & 0.70 & 0.62 & 84.7 \\
\hline
\multicolumn{16}{l}{\textit{e. Image-level supervised methods}} \\
GradCAM$^{*}$ & 0.76$_{0.31}$ & 0.69$_{0.35}$ & 37.4$_{99.2}$ & 0.74$_{0.34}$ & 0.69$_{0.37}$ & 40.0$_{101.2}$ & 0.75$_{0.33}$ & 0.68$_{0.35}$ & 46.5$_{111.0}$ & 0.67$_{0.39}$ & 0.63$_{0.42}$ & 48.6$_{105.5}$ & 0.73 & 0.67 & 43.1 \\
ScoreCAM$^{*}$ & 0.75$_{0.31}$ & 0.69$_{0.35}$ & 37.6$_{99.2}$ & 0.75$_{0.32}$ & 0.70$_{0.36}$ & 38.7$_{101.3}$ & 0.75$_{0.33}$ & 0.69$_{0.35}$ & 46.5$_{111.0}$ & 0.66$_{0.40}$ & 0.62$_{0.43}$ & 50.6$_{105.2}$ & 0.73 & 0.67 & 43.3 \\
LayerCAM$^{*}$ & 0.75$_{0.31}$ & 0.69$_{0.35}$ & 37.5$_{99.2}$ & 0.75$_{0.33}$ & 0.69$_{0.36}$ & 39.4$_{101.3}$ & 0.75$_{0.33}$ & 0.69$_{0.35}$ & 46.5$_{111.0}$ & 0.67$_{0.40}$ & 0.63$_{0.43}$ & 49.9$_{105.5}$ & 0.73 & 0.67 & 43.3 \\
AME-CAM$^{*}$ & 0.73$_{0.33}$ & 0.67$_{0.37}$ & 33.7$_{94.3}$ & 0.75$_{0.31}$ & 0.69$_{0.36}$ & 31.3$_{79.0}$ & 0.69$_{0.35}$ & 0.63$_{0.39}$ & 59.9$_{115.9}$ & 0.76$_{0.32}$ & 0.70$_{0.36}$ & 37.5$_{92.2}$ & 0.73 & 0.67 & 40.6 \\
HarmoFL$^{*}$ & 0.59$_{0.44}$ & 0.57$_{0.47}$ & 74.9$_{106.4}$ & 0.63$_{0.42}$ & 0.60$_{0.45}$ & 63.9$_{91.7}$ & 0.60$_{0.42}$ & 0.56$_{0.45}$ & 84.7$_{119.9}$ & 0.58$_{0.48}$ & 0.57$_{0.49}$ & 82.7$_{106.4}$ & 0.60 & 0.57 & 76.5 \\
FedNorm+$^{*}$ & 0.73$_{0.33}$ & 0.67$_{0.37}$ & 38.3$_{94.9}$ & 0.75$_{0.32}$ & 0.70$_{0.36}$ & 40.1$_{99.4}$ & 0.77$_{0.32}$ & 0.72$_{0.34}$ & 44.6$_{109.9}$ & 0.70$_{0.37}$ & 0.66$_{0.41}$ & 37.9$_{91.0}$ & 0.74 & 0.69 & 40.2 \\
FL-W3S$^{*}$ & 0.73$_{0.35}$ & 0.68$_{0.38}$ & 40.3$_{95.9}$ & 0.81$_{0.29}$ & 0.76$_{0.32}$ & \textbf{26.5}$_{76.8}$ & 0.69$_{0.37}$ & 0.64$_{0.39}$ & 59.2$_{121.2}$ & 0.78$_{0.33}$ & 0.73$_{0.36}$ & \textbf{30.1}$_{81.1}$ & 0.75 & 0.70 & 39.0 \\
\methodname{}$^{*}$ & 0.81$_{0.27}$ & 0.75$_{0.31}$ & 31.8$_{90.6}$ & 0.77$_{0.33}$ & 0.72$_{0.36}$ & 46.1$_{103.4}$ & 0.79$_{0.32}$ & 0.74$_{0.34}$ & 47.1$_{115.0}$ & 0.81$_{0.30}$ & 0.76$_{0.32}$ & 37.6$_{100.5}$ & 0.80 & 0.74 & 40.6 \\
\methodname{} & \textbf{0.84}$_{0.26}$ & \textbf{0.79}$_{0.29}$ & \textbf{28.9}$_{88.2}$ & \textbf{0.86}$_{0.26}$ & \textbf{0.82}$_{0.28}$ & 27.3$_{88.3}$ & \textbf{0.82}$_{0.30}$ & \textbf{0.78}$_{0.31}$ & \textbf{39.1}$_{105.8}$ & \textbf{0.82}$_{0.30}$ & \textbf{0.77}$_{0.32}$ & 35.0$_{98.9}$ & \textbf{0.84} & \textbf{0.79} & \textbf{32.6} \\
\hline
\end{tabular}%
}
\end{table*}

\begin{table*}[t]
\centering
\caption{Segmentation performance on BraTS-MEN per client and averaged across clients. Dice (D, $\uparrow$), IoU (J, $\uparrow$), and HD95 (H, mm, $\downarrow$) are reported; subscripts denote per-slice standard deviation. Fully supervised methods (group~a) and the centralized image-level model (group~b) serve as upper references and are excluded from the comparison. Bold denotes the best federated weakly 
supervised result; $^{*}$ marks baselines \methodname{} significantly outperforms ($p<0.01$, two-sided paired Wilcoxon; all metrics except HD95 for FedDM).}
\label{tab:main_men}
\resizebox{\textwidth}{!}{%
\begin{tabular}{l|ccc|ccc|ccc|ccc|ccc}
\hline
\multirow{2}{*}{Method} 
& \multicolumn{3}{c|}{Client 1} 
& \multicolumn{3}{c|}{Client 2} 
& \multicolumn{3}{c|}{Client 3} 
& \multicolumn{3}{c|}{Client 4} 
& \multicolumn{3}{c}{Average} \\
& D ($\uparrow$) & J ($\uparrow$) & H ($\downarrow$)
& D ($\uparrow$) & J ($\uparrow$) & H ($\downarrow$)
& D ($\uparrow$) & J ($\uparrow$) & H ($\downarrow$)
& D ($\uparrow$) & J ($\uparrow$) & H ( $\downarrow$)
& D ($\uparrow$) & J ($\uparrow$) & H ($\downarrow$) \\
\hline
\multicolumn{16}{l}{\textit{a. Fully supervised federated methods}} \\
FedAvg & 0.76$_{0.40}$ & 0.74$_{0.41}$ & 77.3$_{148.8}$ & 0.86$_{0.33}$ & 0.85$_{0.34}$ & 50.9$_{125.7}$ & 0.84$_{0.34}$ & 0.82$_{0.36}$ & 54.6$_{130.2}$ & 0.79$_{0.38}$ & 0.77$_{0.39}$ & 70.6$_{142.9}$ & 0.81 & 0.79 & 63.3 \\
FedBN & 0.82$_{0.34}$ & 0.80$_{0.36}$ & 48.8$_{119.5}$ & 0.85$_{0.34}$ & 0.83$_{0.35}$ & 52.2$_{128.1}$ & 0.86$_{0.33}$ & 0.84$_{0.34}$ & 49.4$_{125.2}$ & 0.81$_{0.35}$ & 0.79$_{0.37}$ & 59.4$_{132.1}$ & 0.83 & 0.81 & 52.4 \\
FedProx & 0.82$_{0.35}$ & 0.80$_{0.36}$ & 55.6$_{129.6}$ & 0.85$_{0.34}$ & 0.83$_{0.35}$ & 52.7$_{126.6}$ & 0.85$_{0.34}$ & 0.84$_{0.35}$ & 51.5$_{127.9}$ & 0.79$_{0.36}$ & 0.77$_{0.38}$ & 61.3$_{133.3}$ & 0.83 & 0.81 & 55.3 \\
MOON & 0.85$_{0.32}$ & 0.83$_{0.33}$ & 44.3$_{116.4}$ & 0.84$_{0.35}$ & 0.83$_{0.36}$ & 54.1$_{128.6}$ & 0.86$_{0.32}$ & 0.85$_{0.33}$ & 47.5$_{122.9}$ & 0.80$_{0.37}$ & 0.78$_{0.38}$ & 63.9$_{136.6}$ & 0.84 & 0.82 & 52.5 \\
FedDG & 0.86$_{0.31}$ & 0.84$_{0.32}$ & 36.9$_{107.8}$ & 0.85$_{0.35}$ & 0.84$_{0.36}$ & 53.8$_{130.4}$ & 0.83$_{0.35}$ & 0.82$_{0.36}$ & 50.2$_{125.8}$ & 0.74$_{0.41}$ & 0.72$_{0.42}$ & 70.7$_{142.2}$ & 0.82 & 0.81 & 52.9 \\
\hline
\multicolumn{16}{l}{\textit{b. Centralized image-level method (non-federated reference)}} \\
Centralized & 0.78$_{0.36}$ & 0.75$_{0.38}$ & 51.8$_{120.3}$ & 0.88$_{0.29}$ & 0.86$_{0.31}$ & 30.8$_{99.1}$ & 0.83$_{0.34}$ & 0.81$_{0.35}$ & 44.1$_{117.7}$ & 0.78$_{0.36}$ & 0.75$_{0.38}$ & 53.7$_{124.6}$ & 0.82 & 0.79 & 45.1 \\
\hline
\multicolumn{16}{l}{\textit{c. Box-supervised methods}} \\
FedAvg$^{*}$ & 0.77$_{0.38}$ & 0.75$_{0.41}$ & 57.5$_{127.6}$ & 0.81$_{0.37}$ & 0.80$_{0.39}$ & 53.4$_{126.7}$ & 0.78$_{0.40}$ & 0.76$_{0.42}$ & 65.2$_{138.0}$ & 0.73$_{0.41}$ & 0.71$_{0.43}$ & 65.2$_{134.7}$ & 0.77 & 0.76 & 60.3 \\
FedBN$^{*}$ & 0.77$_{0.39}$ & 0.75$_{0.41}$ & 69.3$_{139.5}$ & 0.82$_{0.37}$ & 0.81$_{0.38}$ & 54.4$_{129.1}$ & 0.77$_{0.40}$ & 0.76$_{0.42}$ & 63.1$_{136.0}$ & 0.73$_{0.42}$ & 0.71$_{0.44}$ & 78.8$_{146.8}$ & 0.77 & 0.76 & 66.4 \\
FedProx$^{*}$ & 0.78$_{0.38}$ & 0.75$_{0.40}$ & 52.1$_{122.5}$ & 0.81$_{0.37}$ & 0.80$_{0.39}$ & 56.4$_{130.8}$ & 0.77$_{0.39}$ & 0.76$_{0.41}$ & \textbf{48.2}$_{118.5}$ & 0.72$_{0.42}$ & 0.70$_{0.44}$ & 70.9$_{137.2}$ & 0.77 & 0.75 & 56.9 \\
MOON$^{*}$ & 0.77$_{0.38}$ & 0.75$_{0.41}$ & 56.6$_{126.7}$ & 0.83$_{0.36}$ & 0.82$_{0.37}$ & \textbf{50.1}$_{124.2}$ & 0.76$_{0.40}$ & 0.75$_{0.42}$ & 62.1$_{134.1}$ & 0.72$_{0.41}$ & 0.70$_{0.44}$ & 69.4$_{138.0}$ & 0.77 & 0.76 & 59.5 \\
FedDM$^{*}$ & \textbf{0.80}$_{0.35}$ & \textbf{0.77}$_{0.38}$ & \textbf{38.7}$_{106.2}$ & 0.83$_{0.36}$ & 0.82$_{0.37}$ & 53.7$_{129.2}$ & 0.80$_{0.38}$ & 0.78$_{0.39}$ & 56.0$_{131.3}$ & 0.72$_{0.41}$ & 0.71$_{0.42}$ & 74.7$_{146.5}$ & 0.79 & 0.77 & \textbf{55.8} \\
\hline
\multicolumn{16}{l}{\textit{d. Point-supervised methods}} \\
FedICRA$^{*}$ & 0.64$_{0.42}$ & 0.58$_{0.45}$ & 112.8$_{170.6}$ & 0.67$_{0.41}$ & 0.63$_{0.43}$ & 108.4$_{151.2}$ & 0.75$_{0.41}$ & 0.71$_{0.45}$ & 78.3$_{159.5}$ & 0.65$_{0.41}$ & 0.61$_{0.45}$ & 103.0$_{159.1}$ & 0.68 & 0.63 & 100.6 \\
FedLPPA$^{*}$ & 0.67$_{0.42}$ & 0.63$_{0.44}$ & 102.1$_{160.8}$ & 0.77$_{0.41}$ & 0.76$_{0.42}$ & 85.6$_{156.4}$ & \textbf{0.81}$_{0.36}$ & \textbf{0.79}$_{0.38}$ & 59.1$_{134.7}$ & 0.70$_{0.41}$ & 0.67$_{0.42}$ & 81.9$_{146.3}$ & 0.74 & 0.71 & 82.2 \\
\hline
\multicolumn{16}{l}{\textit{e. Image-level supervised methods}} \\
GradCAM$^{*}$ & 0.60$_{0.45}$ & 0.58$_{0.46}$ & 114.1$_{164.9}$ & 0.82$_{0.37}$ & 0.82$_{0.37}$ & 57.3$_{132.9}$ & 0.78$_{0.39}$ & 0.77$_{0.40}$ & 67.9$_{142.3}$ & 0.75$_{0.39}$ & 0.74$_{0.41}$ & 61.4$_{131.0}$ & 0.74 & 0.72 & 75.2 \\
ScoreCAM$^{*}$ & 0.61$_{0.45}$ & 0.58$_{0.46}$ & 113.0$_{165.2}$ & 0.83$_{0.36}$ & 0.82$_{0.37}$ & 57.0$_{133.0}$ & 0.78$_{0.39}$ & 0.77$_{0.40}$ & 67.9$_{142.3}$ & 0.77$_{0.38}$ & 0.74$_{0.40}$ & 60.6$_{131.1}$ & 0.75 & 0.73 & 74.6 \\
LayerCAM$^{*}$ & 0.60$_{0.45}$ & 0.58$_{0.46}$ & 114.0$_{164.9}$ & 0.82$_{0.37}$ & 0.82$_{0.37}$ & 57.3$_{132.9}$ & 0.78$_{0.39}$ & 0.77$_{0.40}$ & 67.9$_{142.3}$ & 0.76$_{0.39}$ & 0.74$_{0.41}$ & 61.3$_{131.0}$ & 0.74 & 0.72 & 75.2 \\
AME-CAM$^{*}$ & 0.73$_{0.40}$ & 0.70$_{0.42}$ & 71.1$_{137.4}$ & 0.82$_{0.37}$ & 0.81$_{0.38}$ & 60.9$_{136.1}$ & 0.77$_{0.39}$ & 0.76$_{0.40}$ & 70.3$_{143.0}$ & 0.79$_{0.37}$ & 0.76$_{0.38}$ & 61.5$_{133.0}$ & 0.78 & 0.76 & 65.9 \\
HarmoFL$^{*}$ & 0.73$_{0.42}$ & 0.71$_{0.43}$ & 60.6$_{123.2}$ & 0.81$_{0.39}$ & 0.81$_{0.40}$ & 59.0$_{130.9}$ & 0.78$_{0.39}$ & 0.76$_{0.41}$ & 58.6$_{126.0}$ & 0.76$_{0.39}$ & 0.74$_{0.41}$ & 67.0$_{137.0}$ & 0.77 & 0.75 & 61.3 \\
FedNorm+$^{*}$ & 0.76$_{0.39}$ & 0.73$_{0.41}$ & 65.8$_{134.3}$ & 0.80$_{0.38}$ & 0.79$_{0.39}$ & 62.7$_{136.9}$ & 0.80$_{0.37}$ & 0.79$_{0.39}$ & 59.2$_{134.2}$ & 0.76$_{0.40}$ & 0.75$_{0.41}$ & 70.7$_{141.5}$ & 0.78 & 0.76 & 64.6 \\
FL-W3S$^{*}$ & 0.78$_{0.38}$ & 0.76$_{0.39}$ & 58.3$_{128.7}$ & 0.78$_{0.41}$ & 0.77$_{0.41}$ & 68.6$_{139.5}$ & 0.76$_{0.41}$ & 0.75$_{0.42}$ & 69.5$_{139.9}$ & 0.75$_{0.41}$ & 0.73$_{0.42}$ & 80.9$_{149.7}$ & 0.77 & 0.75 & 69.3 \\
\methodname{}$^{*}$ & 0.74$_{0.40}$ & 0.72$_{0.41}$ & 72.5$_{140.9}$ & 0.84$_{0.36}$ & 0.83$_{0.36}$ & 54.4$_{130.2}$ & 0.79$_{0.39}$ & 0.78$_{0.39}$ & 68.8$_{143.8}$ & 0.78$_{0.40}$ & 0.76$_{0.41}$ & 63.5$_{145.4}$ & 0.79 & 0.77 & 64.8 \\
\methodname{} & 0.79$_{0.36}$ & 0.77$_{0.38}$ & 55.5$_{125.0}$ & \textbf{0.84}$_{0.35}$ & \textbf{0.84}$_{0.35}$ & 51.0$_{127.2}$ & 0.80$_{0.38}$ & 0.79$_{0.38}$ & 62.3$_{137.8}$ & \textbf{0.79}$_{0.36}$ & \textbf{0.77}$_{0.38}$ & \textbf{57.4}$_{128.9}$ & \textbf{0.81} & \textbf{0.79} & 56.5 \\
\hline
\end{tabular}%
}
\end{table*}

\begin{table*}[t]
\centering
\caption{Segmentation performance on BraTS-SSA per client and averaged across clients. Dice (D, $\uparrow$), IoU (J, $\uparrow$), and HD95 (H, mm, $\downarrow$) are reported; subscripts denote per-slice standard deviation. Fully supervised methods (group~a) and the centralized image-level model (group~b) serve as upper references and are excluded from the comparison. Bold denotes the best federated weakly 
supervised result; $^{*}$ marks baselines \methodname{} significantly outperforms on all metrics ($p<0.01$, two-sided paired Wilcoxon).}
\label{tab:main_ssa}
\resizebox{\textwidth}{!}{%
\begin{tabular}{l|ccc|ccc|ccc|ccc}
\hline
\multirow{2}{*}{Method} 
& \multicolumn{3}{c|}{Client 1} 
& \multicolumn{3}{c|}{Client 2} 
& \multicolumn{3}{c|}{Client 3} 
& \multicolumn{3}{c}{Average} \\
& D ($\uparrow$) & J ($\uparrow$) & H ( $\downarrow$)
& D ($\uparrow$) & J ($\uparrow$) & H ( $\downarrow$)
& D ($\uparrow$) & J ($\uparrow$) & H ( $\downarrow$)
& D ($\uparrow$) & J ($\uparrow$) & H ( $\downarrow$) \\
\hline
\multicolumn{13}{l}{\textit{a. Fully supervised federated methods}} \\
FedAvg & 0.79$_{0.35}$ & 0.75$_{0.36}$ & 56.8$_{126.5}$ & 0.76$_{0.35}$ & 0.71$_{0.36}$ & 60.3$_{130.2}$ & 0.86$_{0.27}$ & 0.81$_{0.29}$ & 35.9$_{99.3}$ & 0.80 & 0.75 & 51.0 \\
FedBN & 0.82$_{0.31}$ & 0.78$_{0.33}$ & 45.2$_{110.6}$ & 0.77$_{0.34}$ & 0.72$_{0.35}$ & 57.7$_{127.4}$ & 0.87$_{0.26}$ & 0.83$_{0.28}$ & 32.6$_{93.2}$ & 0.82 & 0.77 & 45.2 \\
FedProx & 0.84$_{0.29}$ & 0.80$_{0.31}$ & 36.9$_{100.1}$ & 0.74$_{0.36}$ & 0.69$_{0.37}$ & 62.9$_{128.4}$ & 0.84$_{0.28}$ & 0.79$_{0.31}$ & 34.0$_{93.8}$ & 0.81 & 0.76 & 44.6 \\
MOON & 0.84$_{0.29}$ & 0.80$_{0.31}$ & 38.4$_{102.1}$ & 0.78$_{0.36}$ & 0.74$_{0.36}$ & 66.2$_{137.3}$ & 0.86$_{0.26}$ & 0.81$_{0.28}$ & 36.7$_{90.5}$ & 0.83 & 0.78 & 47.1 \\
FedDG & 0.85$_{0.29}$ & 0.82$_{0.30}$ & 26.7$_{85.0}$ & 0.76$_{0.36}$ & 0.71$_{0.37}$ & 52.8$_{121.6}$ & 0.76$_{0.34}$ & 0.71$_{0.35}$ & 48.9$_{111.8}$ & 0.79 & 0.75 & 42.8 \\
\hline
\multicolumn{13}{l}{\textit{b. Centralized image-level method (non-federated reference)}} \\
Centralized & 0.62$_{0.40}$ & 0.57$_{0.43}$ & 64.5$_{111.7}$ & 0.54$_{0.43}$ & 0.50$_{0.45}$ & 75.8$_{115.1}$ & 0.58$_{0.40}$ & 0.52$_{0.43}$ & 66.0$_{105.7}$ & 0.58 & 0.53 & 68.8 \\
\hline
\multicolumn{13}{l}{\textit{c. Box-supervised methods} (M=20)} \\
FedAvg$^{*}$ & 0.64$_{0.38}$ & 0.58$_{0.42}$ & 59.3$_{113.4}$ & 0.61$_{0.39}$ & 0.56$_{0.43}$ & 65.5$_{122.1}$ & 0.61$_{0.36}$ & 0.56$_{0.41}$ & 49.7$_{92.4}$ & 0.62 & 0.57 & 58.2 \\
FedBN$^{*}$ & 0.58$_{0.41}$ & 0.53$_{0.43}$ & 96.4$_{146.0}$ & 0.60$_{0.40}$ & 0.56$_{0.43}$ & 70.7$_{129.4}$ & 0.61$_{0.36}$ & 0.56$_{0.41}$ & 54.0$_{99.9}$ & 0.60 & 0.55 & 73.7 \\
FedProx$^{*}$ & 0.69$_{0.36}$ & 0.64$_{0.40}$ & 50.1$_{105.8}$ & 0.60$_{0.41}$ & 0.56$_{0.44}$ & 80.7$_{138.8}$ & 0.64$_{0.35}$ & 0.58$_{0.40}$ & 46.1$_{94.2}$ & 0.64 & 0.59 & 59.0 \\
MOON$^{*}$ & 0.64$_{0.38}$ & 0.59$_{0.42}$ & 64.3$_{119.9}$ & 0.60$_{0.40}$ & 0.56$_{0.44}$ & 66.5$_{120.6}$ & 0.65$_{0.35}$ & 0.59$_{0.40}$ & 46.4$_{93.4}$ & 0.63 & 0.58 & 59.1 \\
FedDM$^{*}$ & 0.75$_{0.33}$ & 0.69$_{0.36}$ & \textbf{41.4}$_{105.8}$ & 0.62$_{0.42}$ & 0.58$_{0.44}$ & 89.3$_{151.0}$ & 0.69$_{0.32}$ & 0.62$_{0.37}$ & 39.8$_{94.3}$ & 0.69 & 0.63 & 56.8 \\
\hline
\multicolumn{13}{l}{\textit{d. Point-supervised methods}} \\
FedICRA$^{*}$ & 0.67$_{0.39}$ & 0.60$_{0.39}$ & 93.4$_{152.3}$ & 0.59$_{0.40}$ & 0.53$_{0.41}$ & 104.5$_{157.9}$ & 0.66$_{0.39}$ & 0.59$_{0.39}$ & 86.6$_{142.8}$ & 0.64 & 0.57 & 94.8 \\
FedLPPA$^{*}$ & 0.63$_{0.41}$ & 0.58$_{0.41}$ & 104.2$_{156.5}$ & 0.70$_{0.37}$ & 0.64$_{0.39}$ & 64.7$_{129.5}$ & 0.77$_{0.32}$ & 0.71$_{0.35}$ & 47.0$_{110.8}$ & 0.70 & 0.64 & 72.0 \\
\hline
\multicolumn{13}{l}{\textit{e. Image-level supervised methods}} \\
GradCAM$^{*}$ & 0.61$_{0.39}$ & 0.55$_{0.41}$ & 85.8$_{142.7}$ & 0.71$_{0.35}$ & 0.65$_{0.37}$ & 56.7$_{122.3}$ & 0.71$_{0.33}$ & 0.64$_{0.36}$ & 50.8$_{111.1}$ & 0.68 & 0.61 & 64.4 \\
ScoreCAM$^{*}$ & 0.64$_{0.39}$ & 0.58$_{0.40}$ & 83.5$_{143.4}$ & 0.71$_{0.35}$ & 0.65$_{0.37}$ & \textbf{56.4}$_{122.4}$ & 0.74$_{0.32}$ & 0.67$_{0.34}$ & 48.4$_{111.6}$ & 0.70 & 0.63 & 62.8 \\
LayerCAM$^{*}$ & 0.62$_{0.39}$ & 0.56$_{0.40}$ & 85.3$_{142.8}$ & 0.71$_{0.35}$ & 0.65$_{0.37}$ & 56.5$_{122.4}$ & 0.71$_{0.33}$ & 0.64$_{0.36}$ & 50.8$_{111.1}$ & 0.68 & 0.62 & 64.2 \\
AME-CAM$^{*}$ & 0.62$_{0.40}$ & 0.57$_{0.43}$ & 78.6$_{128.6}$ & 0.65$_{0.40}$ & 0.61$_{0.41}$ & 84.0$_{145.4}$ & 0.58$_{0.39}$ & 0.52$_{0.43}$ & 59.8$_{106.1}$ & 0.62 & 0.57 & 74.1 \\
HarmoFL$^{*}$ & 0.64$_{0.41}$ & 0.61$_{0.44}$ & 61.9$_{106.8}$ & 0.68$_{0.37}$ & 0.62$_{0.39}$ & 75.2$_{127.5}$ & 0.49$_{0.47}$ & 0.47$_{0.48}$ & 63.5$_{96.8}$ & 0.60 & 0.57 & 66.8 \\
FedNorm+$^{*}$ & 0.69$_{0.37}$ & 0.64$_{0.40}$ & 57.7$_{117.0}$ & 0.69$_{0.36}$ & 0.63$_{0.39}$ & 50.6$_{113.0}$ & 0.69$_{0.35}$ & 0.62$_{0.38}$ & 60.4$_{119.2}$ & 0.69 & 0.63 & 56.2 \\
FL-W3S$^{*}$ & 0.67$_{0.38}$ & 0.61$_{0.42}$ & 54.8$_{102.8}$ & 0.58$_{0.41}$ & 0.54$_{0.44}$ & 62.1$_{104.2}$ & 0.63$_{0.37}$ & 0.56$_{0.41}$ & 56.0$_{98.1}$ & 0.63 & 0.57 & 57.7 \\
\methodname{}$^{*}$ & 0.78$_{0.32}$ & 0.73$_{0.34}$ & 45.4$_{109.9}$ & 0.66$_{0.42}$ & 0.63$_{0.43}$ & 102.4$_{161.2}$ & 0.80$_{0.29}$ & 0.74$_{0.32}$ & \textbf{38.1}$_{96.0}$ & 0.75 & 0.70 & 62.0 \\
\methodname{} & \textbf{0.78}$_{0.34}$ & \textbf{0.74}$_{0.35}$ & 53.7$_{119.8}$ & \textbf{0.75}$_{0.35}$ & \textbf{0.70}$_{0.36}$ & 61.1$_{130.5}$ & \textbf{0.82}$_{0.29}$ & \textbf{0.77}$_{0.30}$ & 39.9$_{104.8}$ & \textbf{0.78} & \textbf{0.73} & \textbf{51.6} \\
\hline
\end{tabular}%
}
\end{table*}

\section{Results}
\label{subsec:results}
\subsection{Comparison with state-of-the-art methods}
\subsubsection{FeTS2022}
Table~\ref{tab:main} reports per-client and averaged segmentation 
performance on the FeTS2022 dataset. Among all weakly supervised methods, MOSAIC consistently achieves the best Dice and IoU across every client, reaching an average Dice of 0.84 and IoU of 0.79 using only 
image-level labels. All reported improvements over competing methods 
are statistically significant under the two-sided paired Wilcoxon signed-rank test ($p < 0.01$; starred rows in Table~\ref{tab:main}).

Compared with the image-level CAM baselines (GradCAM, ScoreCAM, LayerCAM), which plateau at approximately 0.73 Dice and 0.67 IoU, MOSAIC improves Dice by 0.11 and IoU by 0.12, corresponding to relative gains of 15\% and 18\%, while reducing HD95 from the 42--43\,mm range to 32.6\,mm. The improvement is consistent across all clients, demonstrating that the modality-agnostic fusion representation generalizes across heterogeneous modality subsets rather than benefiting a single privileged site. Most notably, on the FLAIR-only Client~4, the most modality-scarce configuration, MOSAIC achieves 0.82 Dice compared with 0.58--0.78 for existing image-level approaches, confirming that SPA effectively compensates for severely incomplete multimodal fusion inputs by aligning the projected representation to a global spectral consensus rather than relying on absent channel statistics.

MOSAIC surpasses methods relying on substantially richer supervision. The best box-supervised baseline (FedDM) achieves 0.75 Dice, while point-supervised FedICRA and FedLPPA obtain only 0.72 and 0.70 Dice respectively, with substantially larger boundary errors. Among the image-level federated baselines, FL-W3S achieves 0.75 Dice, the strongest competing image-level method, yet remains 0.09 Dice below MOSAIC, with a relative gap of 12\%. HarmoFL achieves only 0.60 Dice with HD95 of 76.5\,mm, demonstrating that directly harmonizing Fourier amplitude at the raw input level is insufficient under heterogeneous missing modalities: the absent channels corrupt the input amplitude distribution before any alignment can occur, whereas SPA operates in feature space after modality projection and aligns compact spectral statistics of the fused representation. FedNorm+, the only baseline explicitly designed for missing modalities, reaches 0.74 Dice and 0.69 IoU under image-level supervision, with a relative gap of 14\% compared to MOSAIC. Its modality-gated normalization compensates for absent channels less effectively than SPA, and degrades most on the single-modality Client~4 (0.70 Dice), precisely the configuration where incomplete multimodal fusion is most severe. Among the fully supervised references, FedDG attains 0.85 Dice, below the other dense-label baselines (0.87--0.88), indicating that interpolating raw amplitude spectra across clients is less suited to heterogeneous modality subsets than learning modality-agnostic representations.

\begin{figure*}[!t]
\centering
\includegraphics[width=0.8\textwidth]{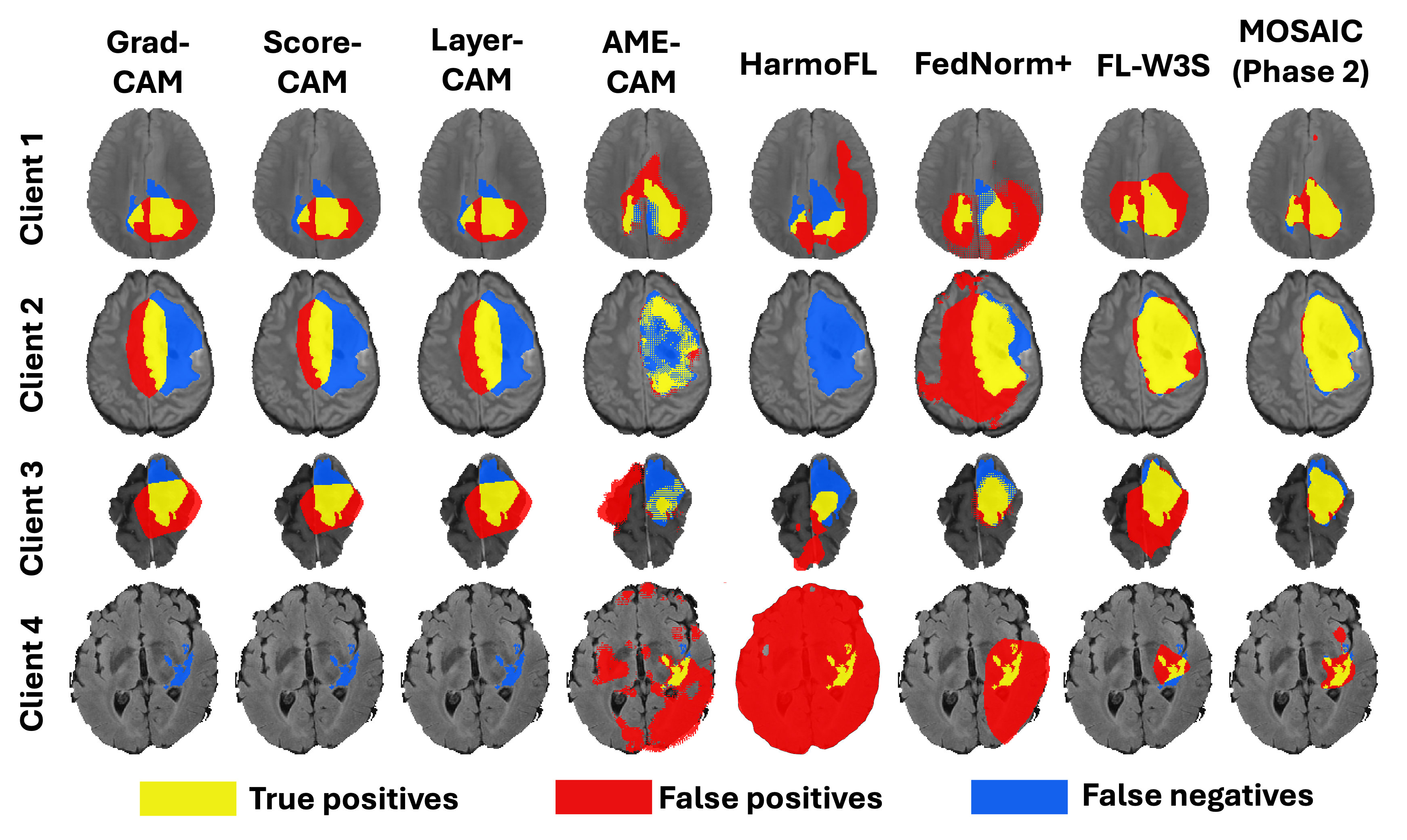}
\caption{Visual comparison of MOSAIC with state-of-the-art methods on FeTS2022. A representative test slice for each client shown with predictions overlaid on the input slice. True positives, false positives, and false negatives are shown in yellow red and blue respectively.}
\label{fig:qual_fets}
\end{figure*}

Despite operating under strict privacy constraints and client-specific 
missing modalities, MOSAIC approaches the performance of fully supervised federated methods, reducing the gap to less than 0.05 Dice while requiring no dense pixel annotations. Strikingly, MOSAIC also surpasses the centralized image-level reference, a non-federated model trained on pooled data with the complete modality set, by 0.02 Dice (0.84 vs. 0.82, a 2\% relative gain) and 0.03 IoU (4\% relative), while maintaining comparable boundary quality (32.6 vs. 29.9\,mm HD95). This result demonstrates a key property of MOSAIC: federated multimodal fusion with client-specific adaptation can match or exceed the performance achievable by centralizing data, because the federation enforces site-specific alignment rather than averaging across heterogeneous inputs. Figure~\ref{fig:qual_fets} provides a qualitative comparison, where MOSAIC produces the largest true-positive coverage with markedly fewer false positives than box-supervised baselines and fewer false negatives than CAM baselines across all four modality subsets.

\subsubsection{Generalization to BraTS-MEN}
Table~\ref{tab:main_men} reports the same comparison on BraTS-MEN, which differs from FeTS2022 in tumor type (meningioma vs. glioma), clinical population, and client-wise modality assignment. All improvements over competing methods are statistically significant on Dice and IoU, and on HD95 for every baseline except FedDM ($p < 0.01$). The trends observed on FeTS2022 carry over consistently, confirming that MOSAIC's multimodal fusion alignment strategy generalizes across tumor types and modality configurations.

MOSAIC attains the best average Dice (0.81) and IoU (0.79), 
improving on the image-level CAM baseline (AME-CAM at 0.78 Dice) 
by 0.03 Dice (4\% relative) and on box-supervised FedDM (0.79) by 
0.02 Dice (3\% relative) using only image-level labels. The refinement stage raises Phase~1 pseudo-labels from 0.79 to 0.81 Dice while reducing average HD95 from 64.8 to 56.5\,mm. Robustness to modality scarcity is most 
evident on the single-modality Client~4, where MOSAIC reaches 0.79 
Dice, the best among all weakly supervised methods and higher than 
every box and point-supervised baseline on that client.

\begin{figure*}[!t]
\centering
\includegraphics[width=0.8\textwidth]{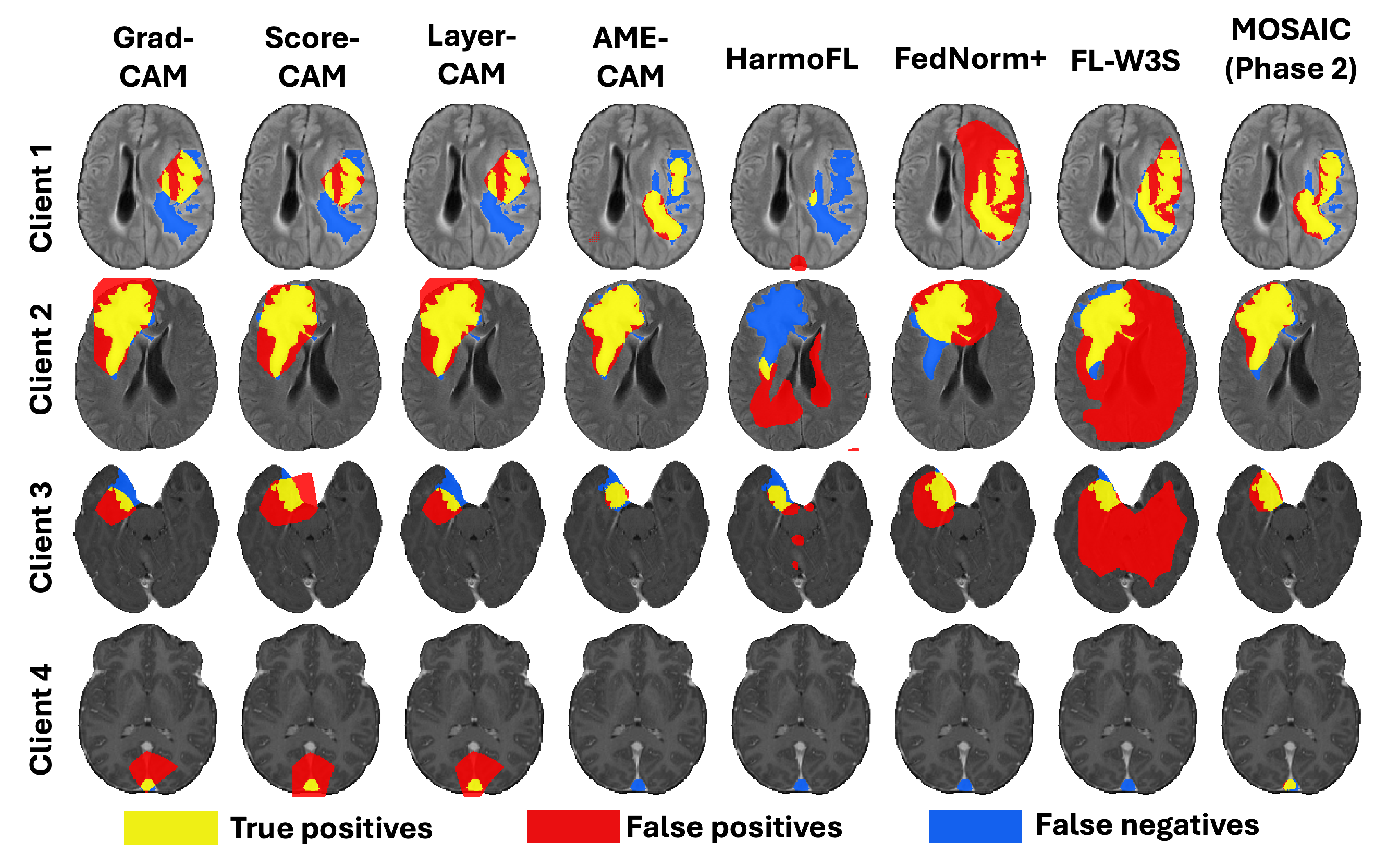}
\caption{Visual comparison of methods on BraTS-MEN. A representative test slice for each client shown with predictions overlaid on the input slice. True positives, false positives, and false negatives are shown in yellow red and blue respectively.}
\label{fig:qual_men}
\end{figure*}

Among additional image-level baselines, HarmoFL is more competitive on BraTS-MEN than FeTS2022, achieving 0.77 Dice with the lowest boundary error among image-level baselines (61.3\, mm), yet 
remaining 0.04 Dice below MOSAIC (5\% relative). FL-W3S achieves 
0.77 Dice with a narrow 0.04 client spread, but remains 0.04 Dice 
below our full model (5\% relative) with higher boundary error 
(69.3 vs. 56.5\,mm). FedNorm+ reaches 0.78 Dice and 0.76 IoU on this 
dataset, the second-best image-level result, yet still trails 
MOSAIC by 0.03 Dice (4\% relative) with considerably larger boundary
error (64.6 vs. 56.5\,mm). Its relative improvement on BraTS-MEN compared to FeTS2022 reflects that the meningioma dataset has a more balanced modality distribution across clients, reducing the severity of incomplete fusion that MOSAIC is specifically designed to address. FedDG reaches 0.82 Dice with dense supervision; our image-level 
model is within 0.01 Dice and surpasses it on the T1ce-only 
Client~4 (0.79 vs. 0.74), where amplitude interpolation across 
mismatched modality subsets is least reliable. The centralized 
image-level reference reaches 0.82 Dice and 0.79 IoU; our 
federated model remains within 0.01 Dice of this ceiling 
despite never sharing data or complete modality sets, demonstrating 
near-lossless privacy-preserving multimodal fusion on this dataset. The qualitative comparison (Figure~\ref{fig:qual_men}) highlights robustness on Client~4, where competing methods largely fail on a small T1ce-only lesion that MOSAIC successfully recovers, demonstrating the value of the global prototype cue in providing a CAM-independent fusion signal when a single discriminative modality is absent.

\subsubsection{Generalization to BraTS-SSA}
Table~\ref{tab:main_ssa} reports the same comparison on BraTS-SSA, a smaller three-client Sub-Saharan African glioma cohort that stresses the framework under limited data, an underrepresented clinical population, and a three-client federation structure. Every comparison against weakly supervised baselines pooled over the three clients is statistically significant on all three metrics ($p < 0.01$; Table~\ref{tab:main_ssa}). MOSAIC attains the best average Dice (0.78) and IoU (0.73) while achieving the lowest average boundary error (51.6\,mm), narrows the gap to the fully supervised upper reference ($\approx0.80$--$0.83$ average Dice) and improving on the best image-level CAM baseline (ScoreCAM at 0.70 Dice) by 0.08 Dice (11\% relative) and on the best box-supervised method (FedDM at 0.69) by 0.09 Dice (13\% relative) using only image-level labels. Our full model is the best weakly supervised method on every client for both Dice and IoU, including the FLAIR-only Client~1 ($78.1\%$ Dice), and  to under $4$ Dice points. 

\begin{figure*}[!t]
\centering
\includegraphics[width=0.8\textwidth]{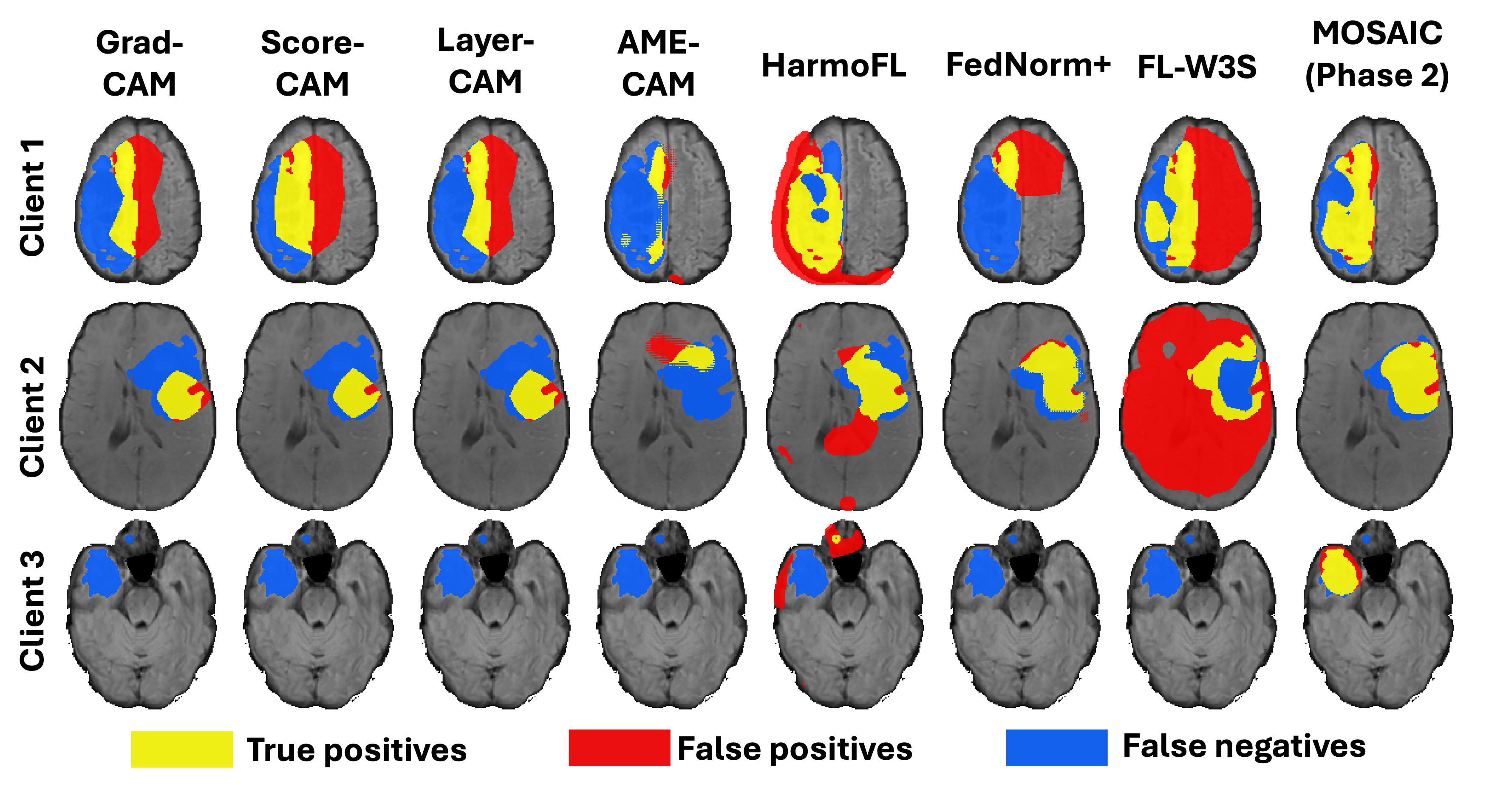}
\caption{Visual comparison of methods on BraTS-SSA. A representative test slice for each client shown with predictions overlaid on the input slice. True positives, false positives, and false negatives are shown in yellow red and blue respectively. }
\label{fig:qual_ssa}
\end{figure*}

The refinement stage is decisive: Phase~1 pseudo-labels improve from 0.75 to 0.78 average Dice, with the largest gain on the \{T1ce,~T2\} Client~2 (0.66$\rightarrow$0.75, HD95 from 102.4 to 61.1\,mm) where the absent FLAIR modality causes severe CAM mislocalization, and the prototype cue provides the critical CAM-independent fusion signal to recover reliable supervision. FedNorm+ is the image-level baseline on this cohort at 0.69 Dice, but remains 0.09 Dice below MOSAIC (13\% relative), confirming that modality-gated normalization alone cannot substitute for the spectral alignment of multimodal fusion representations across clients. FedDG, trained with dense masks, obtains 0.79 Dice; our image-level model matches this to within 0.01 Dice despite requiring no voxel annotations. The limited-data setting further exposes differences among baselines: HarmoFL achieves only 0.60 Dice and FL-W3S reaches 0.63, both substantially below our model and the CAM baselines. Notably, the centralized reference performs substantially worse at 0.58 Dice, suggesting that pooling the three small client cohorts does not provide sufficient data diversity for 
robust centralized CAM fusion, a regime where MOSAIC's federated client-specific adaptation is advantageous over naive data pooling. Figure~\ref{fig:qual_ssa} qualitatively confirms that MOSAIC attains the closest agreement with ground truth across all three SSA clients.

\subsubsection{Cross-dataset consistency}
Across two tumor types, three clinical populations, two federation sizes, and three structurally distinct modality configurations, MOSAIC achieves consistent and statistically significant superiority on all datasets, confirming that modality-agnostic spectral alignment is a general principle rather than a dataset-specific artifact. The persistent gap between MOSAIC and FedNorm+, the only competing method designed for missing modalities, across all three datasets further confirms that frequency-domain feature alignment provides a qualitatively more effective signal than modality-gated normalization for heterogeneous multimodal fusion under privacy constraints.

\subsubsection{Effect of the refinement stage}
The two-phase design separates modality-robust pseudo-label generation (Phase~1) from refinement-based denoising (Phase~2). Phase~1 alone achieves $0.80$ average Dice on FeTS2022, 0.79 on BraTS-MEN, and 0.75 on BraTS-SSA, outperforming every competing weakly supervised method on all three datasets. This demonstrates that the modality-agnostic multimodal fusion alignment is the primary driver of MOSAIC's performance advantage. Phase~2 adds consistent Dice gains of
$+0.04$ ($5\%$ gain), $+0.02$ ($3\%$ gain), and $+0.03$ ($4\%$ gain) respectively across the three datasets, with the largest individual gains where missing modalities cause the most severe CAM mislocalization (Client~2, FeTS2022: 0.77$\rightarrow$0.86; Client~2, BraTS-SSA: 
0.66$\rightarrow$0.75). These gains confirm that the soft-vote arbitration and trimap denoising effectively recover accurate boundaries from noisy pseudo-labels, while the global prototype cue provides a CAM-independent fusion signal that rescues modality-poor clients whose local pseudo-labels are systematically unreliable. The two phases are complementary: Phase~1 generates robust fused pseudo-labels under missing modalities, Phase~2 refines them into accurate segmentation masks.

\subsection{Ablation on Spectral Prototype Resolution}
Table~\ref{tab:ablation_bands} reports the effect of radial band count 
$\Omega$ on FeTS2022. Performance improves steadily from $\Omega=2$ 
(0.81 Dice) to $\Omega=8$ (0.84 Dice, 0.79 IoU), then 
degrades for larger values, dropping sharply at $\Omega=16$ to 
0.79 Dice. The non-monotonic behavior at large $\Omega$ reflects 
overfitting of the spectral prototype to client-specific 
high-frequency noise rather than modality-induced distributional 
signatures: finer bands capture texture detail but less transferable 
style information, reducing the generalization of the alignment 
target across heterogeneous fusion inputs. The boundary error follows 
a related trend, reaching its lowest values around $\Omega=12$--14 
(31.8--32.0\,mm), only marginally below the 32.6\,mm at 
$\Omega=8$, confirming that $\Omega=8$ provides the best 
overlap-boundary trade-off. Importantly, performance remains within 
0.025 Dice points of the optimum across all tested values, 
demonstrating that MOSAIC is robust to the choice of spectral 
resolution and does not require fine-grained hyperparameter tuning.

\subsection{Comparison of feature-alignment objectives}
\label{subsec:results_alignment_obj}
Table~\ref{tab:alignment_obj} compares SPA against four alternative privacy-preserving alignment objectives, each substituted as a drop-in replacement for the SPA loss during Phase~1. SPA achieves the best result on every metric (0.80 Dice, 0.74 IoU, 40.6 mm HD95), with all pairwise differences statistically significant ($p < 0.01$). Among spatial-statistic baselines, mean alignment reaches 0.73 Dice while standard-deviation alignment is markedly weaker at 0.65 Dice with the largest boundary error (67.5 mm), 
indicating that low-order channel moments are insufficient to capture 
the modality-specific textural and structural signatures that distinguish clients operating under heterogeneous fusion inputs. The contrastive InfoNCE objective, which aligns pixel features to shared prototypes, improves over the standard-deviation baseline at 0.70 Dice but still trails SPA by 0.1 Dice. The histogram-KL objective, which matches the full intensity distribution, is the strongest alternative at 0.78 Dice, yet remains 0.02 Dice ($3\%$) and $0.6$\,mm HD95 below SPA. These results establish a clear ordering that validates the core design principle of MOSAIC: aligning in the frequency domain, where modality-specific textural signatures are compactly summarized by radial band energies, is more effective than matching spatial-domain moments, full intensity histograms, or prototype assignments for cross-client multimodal fusion alignment 
under heterogeneous and incomplete modality subsets. Critically, the privacy properties of all five approaches are equivalent under the shared EMA communication protocol, confirming that SPA's advantage derives from the quality of the alignment signal rather than any difference in communication overhead or privacy constraints.

\begin{table}[!t]
\centering
\caption{Ablation on the number of radial spectral bands $\Omega$ in the SPA loss, on FeTS2022. Macro-averaged Dice ($\uparrow$), IoU ($\uparrow$) and HD95 (mm, $\downarrow$). $\Omega=8$ is our default configuration; bold denotes the best value per column.}
\label{tab:ablation_bands}
\begin{tabular}{cccc}
\hline
$\Omega$ (bands) & Dice & IoU & HD95 \\
\hline
2  & 0.81$_{0.29}$ & 0.76$_{0.32}$ & 36.3$_{100.6}$ \\
4  & 0.82$_{0.29}$ & 0.76$_{0.31}$ & 34.7$_{100.1}$ \\
6  & 0.82$_{0.29}$ & 0.77$_{0.31}$ & 35.0$_{99.0}$ \\
8  & \textbf{0.84}$_{0.28}$ & \textbf{0.79}$_{0.30}$ & 32.6$_{95.1}$ \\
10 & 0.81$_{0.29}$ & 0.76$_{0.32}$ & 34.8$_{97.4}$ \\
12 & 0.82$_{0.28}$ & 0.77$_{0.31}$ & 32.0$_{93.8}$ \\
14 & 0.82$_{0.28}$ & 0.77$_{0.31}$ & \textbf{31.8}$_{93.5}$ \\
16 & 0.79$_{0.30}$ & 0.74$_{0.33}$ & 38.6$_{100.5}$ \\
\hline
\end{tabular}
\end{table}

\begin{table}[!t]
\centering
\caption{Feature-alignment objectives compared as drop-in replacements for SPA during Phase~1 on FeTS2022. Macro-averaged Dice ($\uparrow$), IoU ($\uparrow$), and HD95 (mm, $\downarrow$) across four clients; $^{*}$ denotes objectives SPA significantly outperforms on all metrics ($p<0.01$, two-sided paired Wilcoxon).}
\label{tab:alignment_obj}
\begin{tabular}{lccc}
\hline
Alignment objective & Dice & IoU & HD95 \\
\hline
Std$^{*}$            & 0.65 & 0.61 & 67.5 \\
Contrastive$^{*}$     & 0.70 & 0.66 & 51.6 \\
Mean$^{*}$           & 0.73 & 0.67 & 58.2 \\
Histogram KL$^{*}$   & 0.78 & 0.73 & 41.2 \\
SPA (ours)     & \textbf{0.80} & \textbf{0.74} & \textbf{40.6} \\
\hline
\end{tabular}
\end{table}

\subsection{Ablation on model and loss components}
\label{subsec:results_components}
Table~\ref{tab:ablation_components} reports the leave-one-out component 
ablation on FeTS2022. Every component contributes positively, confirming that the Phase~2 objective is well-balanced. The presence classification term $\mathcal{L}_{\mathrm{cls}}$ is the most critical: its removal drops Dice from 0.84 to 0.80 and raises HD95 from 32.6\,mm to 39.1\,mm, as the model loses the slice-level gate that suppresses false positives on tumor-free slices, a failure mode directly amplified by missing modalities that reduce the discriminability of the fused representation on negative slices. 
Trimap denoising is the next most influential ($-0.02$ Dice, $+3.0$\,mm HD95): without it, the network fits the noisy CAM contour, degrading both region overlap and boundary accuracy. The 2.5D context and teacher--student consistency each contribute approximately 0.01 Dice. The core-only variant (modality-alignment module $+$ SPA only) already achieves 0.82 
Dice, surpassing every competing weakly supervised method (best 0.75); the full refinement pipeline adds a further 0.02 Dice and reduces HD95 from 39.6\,mm to 32.6\,mm. This confirms that the two core novel components, modality-agnostic fusion alignment and spectral prototype regularization, account for the bulk of MOSAIC's improvement, while the refinement components sharpen boundary accuracy and suppress pseudo-label noise.

\begin{table}[!t]
\centering
\caption{Leave-one-out model and loss component ablation on FeTS2022. Macro-averaged Dice ($\uparrow$), IoU ($\uparrow$), and HD95 (mm, $\downarrow$) over four clients. Each row disables one component while keeping all others fixed; core-only retains only the modality-alignment module and SPA. Bold denotes the best value per column.}
\label{tab:ablation_components}
\resizebox{\columnwidth}{!}{%
\begin{tabular}{lccc}
\hline
Configuration & Dice & IoU & HD95 \\
\hline
Full \methodname{} & \textbf{0.84}$_{0.28}$ & \textbf{0.79}$_{0.30}$ & 32.6$_{95.1}$ \\
\hline
\quad w/o trimap denoising & 0.82$_{0.29}$ & 0.78$_{0.31}$ & 35.6$_{99.5}$ \\
\quad w/o CRF loss $\mathcal{L}_{\mathrm{CRF}}$ & 0.83$_{0.29}$ & 0.78$_{0.31}$ & 33.8$_{96.4}$ \\
\quad w/o prototype alignment $\mathcal{L}_{\mathrm{proto}}$ & 0.83$_{0.28}$ & 0.78$_{0.31}$ & 33.1$_{96.4}$ \\
\quad w/o consistency $\mathcal{L}_{\mathrm{cons}}$ & 0.83$_{0.29}$ & 0.78$_{0.31}$ & 34.8$_{98.4}$ \\
\quad w/o presence $\mathcal{L}_{\mathrm{cls}}$ & 0.80$_{0.31}$ & 0.75$_{0.33}$ & 39.1$_{102.1}$ \\
\quad w/o 2.5D context ($K{=}1$) & 0.83$_{0.29}$ & 0.78$_{0.31}$ & 34.1$_{97.6}$ \\
\quad w/o soft-vote arbitration & 0.83$_{0.28}$ & 0.78$_{0.30}$ & 32.6$_{95.0}$ \\
\quad w/o RCSA $\mathcal{L}_{\mathrm{RCSA}}$ & 0.83$_{0.28}$ & 0.78$_{0.31}$ & \textbf{31.7}$_{93.6}$ \\
\hline
Core only (adapter $+$ SPA) & 0.82$_{0.30}$ & 0.77$_{0.32}$ & 39.6$_{105.2}$ \\
\hline
\end{tabular}%
}
\end{table}

\subsection{Dynamic Client Addition}
\label{subsec:results_newclient}
Table~\ref{tab:newclient} evaluates MOSAIC's ability to incorporate a 
previously unseen fifth client after the original four-client federation has been trained. Across all four modality configurations, the new client consistently achieves 0.80--0.82 Dice, remaining within only 0.01--0.04 Dice values of the jointly trained base clients (0.82--0.86), confirming genuine open federation. This demonstrates that the count-based modality-alignment module enables new institutions with previously unseen modality combinations to join without retraining or modality bookkeeping, which is critical for real-world multi-centre deployments where acquisition protocols and participating institutions evolve over time.

Both joining strategies are effective. Local-adapt (LA) provides the 
most consistent improvements across configurations: for the \{FLAIR,~T1ce\} subset, the warm-up stage increases Dice from 0.78 to 0.81 while reducing HD95 from 32.3 to 21.5 mm, demonstrating the benefit of brief client-specific modality adapter alignment before full federation. For the FLAIR-only configuration, direct-join (DJ) slightly outperforms, achieving the highest Dice (0.82) and lowest boundary error (20.5 mm), showing that even 
the most modality-scarce clients can participate immediately without 
adaptation overhead. For all other modality subsets the performance 
difference between LA and DJ remains below 0.01 Dice, indicating that the pretrained global backbone, learned from aligned multimodal fusion representations across the original four clients, transfers effectively to unseen institutions regardless of their modality configuration.

\begin{table}[!t]
\centering
\caption{Dynamic client addition on FeTS2022. Segmentation performance of the newly added fifth client, under both local-adapt (LA) and direct-join (DJ) joining strategies. The better of the two strategies for each metric indicated in bold.}
\label{tab:newclient}
\resizebox{\columnwidth}{!}{%
\begin{tabular}{lccc|ccc}
\hline
 & \multicolumn{3}{c|}{Local-adapt (LA)} & \multicolumn{3}{c}{Direct-join (DJ)} \\
Mod. & Dice & IoU & HD95 & Dice & IoU & HD95 \\
\hline
F, T1ce & \textbf{0.81}$_{0.27}$ & \textbf{0.76}$_{0.31}$ & \textbf{21.5}$_{65.1}$ & 0.78$_{0.30}$ & 0.72$_{0.33}$ & 32.3$_{87.0}$ \\
F, T2 & \textbf{0.80}$_{0.28}$ & 0.74$_{0.32}$ & \textbf{22.0}$_{67.0}$ & 0.80$_{0.29}$ & 0.74$_{0.32}$ & 29.1$_{82.6}$ \\
T1ce, T2 & \textbf{0.81}$_{0.30}$ & 0.76$_{0.32}$ & \textbf{37.6}$_{100.9}$ & 0.81$_{0.31}$ & 0.76$_{0.32}$ & 38.9$_{102.2}$ \\
F & 0.81$_{0.27}$ & 0.75$_{0.31}$ & 22.0$_{67.0}$ & \textbf{0.82}$_{0.27}$ & \textbf{0.76}$_{0.31}$ & \textbf{20.5}$_{64.0}$ \\
\hline
\end{tabular}%
}
\end{table}

\subsection{Aleatoric Uncertainty Analysis}
\label{subsec:results_uncertainty}
\begin{figure}[!t]
\centering
\includegraphics[width=\columnwidth]{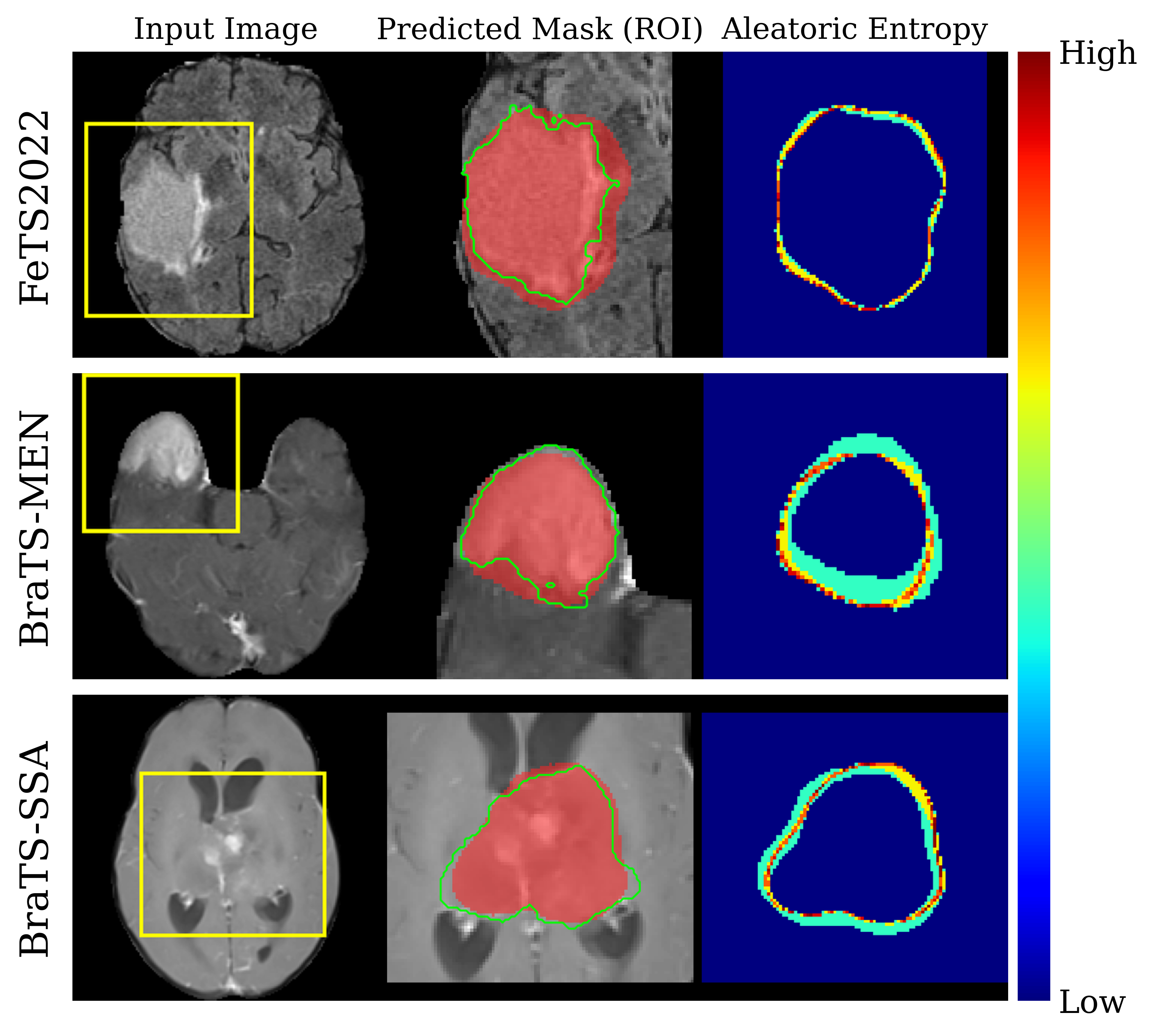}
\caption{Qualitative aleatoric uncertainty for a representative slice from each dataset. Each row shows the input image with the region of interest, the predicted tumor mask (red) with its ground truth contour (green), and the pixel-wise aleatoric entropy.}
\label{fig:unc_panels}
\end{figure}

Table~\ref{tab:uncertainty} reports aleatoric uncertainty under TTA across all three datasets. A consistent and clinically meaningful pattern emerges: foreground entropy $H_{\mathrm{fg}}$ is concentrated at tumor boundaries while background entropy $H_{\mathrm{bg}}$ remains low (average 0.0047, 0.0033, and 0.0039 on FeTS2022, BraTS-MEN, and BraTS-SSA 
respectively), and segmentation stability $\mathrm{Dice}_{\mathrm{unc}}$ 
remains below 0.035 on all datasets and clients. The highest foreground entropy is consistently observed for the \{T1ce,~T2\} modality subset across datasets, reflecting the absence of FLAIR, the sequence providing strongest tumor boundary contrast, and confirming that the model's uncertainty correctly tracks the information content of the available fused modalities. This pattern is clinically desirable from a trustworthy multimodal fusion perspective: uncertainty concentrates precisely where the fused representation is inherently ambiguous due to absent modalities or 
anatomical boundary ambiguity, while the model remains confident where fusion is reliable. Figure~\ref{fig:unc_panels} qualitatively confirms this across all three datasets, with uncertainty confined to a narrow boundary band regardless of tumor type or modality configuration, demonstrating stable and trustworthy fusion behavior under heterogeneous and incomplete multimodal inputs.

\begin{table}[!t]
\centering
\caption{Aleatoric uncertainty under TTA on three datasets. $H_{\mathrm{fg}}$ and $H_{\mathrm{bg}}$: mean predictive entropy over foreground and background pixels; $\mathrm{Dice}_{\mathrm{unc}}$: standard deviation of per-pass Dice under input perturbations. Lower values indicate more confident, stable predictions.}
\label{tab:uncertainty}
\begin{tabular}{llccc}
\hline
Dataset & Client & $H_{\mathrm{fg}}$ & $H_{\mathrm{bg}}$ & $\mathrm{Dice}_{\mathrm{unc}}$ \\
\hline
\multirow{5}{*}{FeTS2022} & 1 & 0.0500 & 0.0042 & 0.0164 \\
 & 2 & 0.0527 & 0.0041 & 0.0207 \\
 & 3 & 0.0729 & 0.0065 & 0.0229 \\
 & 4 & 0.0615 & 0.0040 & 0.0200 \\
 & Avg. & 0.0587 & 0.0047 & 0.0198 \\
\hline
\multirow{5}{*}{BraTS-MEN} & 1 & 0.0775 & 0.0030 & 0.0318 \\
 & 2 & 0.0777 & 0.0031 & 0.0226 \\
 & 3 & 0.1427 & 0.0059 & 0.0415 \\
 & 4 & 0.0759 & 0.0014 & 0.0385 \\
 & Avg. & 0.0927 & 0.0033 & 0.0344 \\
\hline
\multirow{4}{*}{BraTS-SSA} & 1 & 0.0587 & 0.0044 & 0.0152 \\
 & 2 & 0.1077 & 0.0028 & 0.0385 \\
 & 3 & 0.0560 & 0.0046 & 0.0177 \\
 & Avg. & 0.0741 & 0.0039 & 0.0239 \\
\hline
\end{tabular}
\end{table}

\subsection{Analysis of Modality Alignment}
\label{subsec:results_alignment}
\begin{figure}[!t]
\centering
\includegraphics[width=0.45\textwidth]{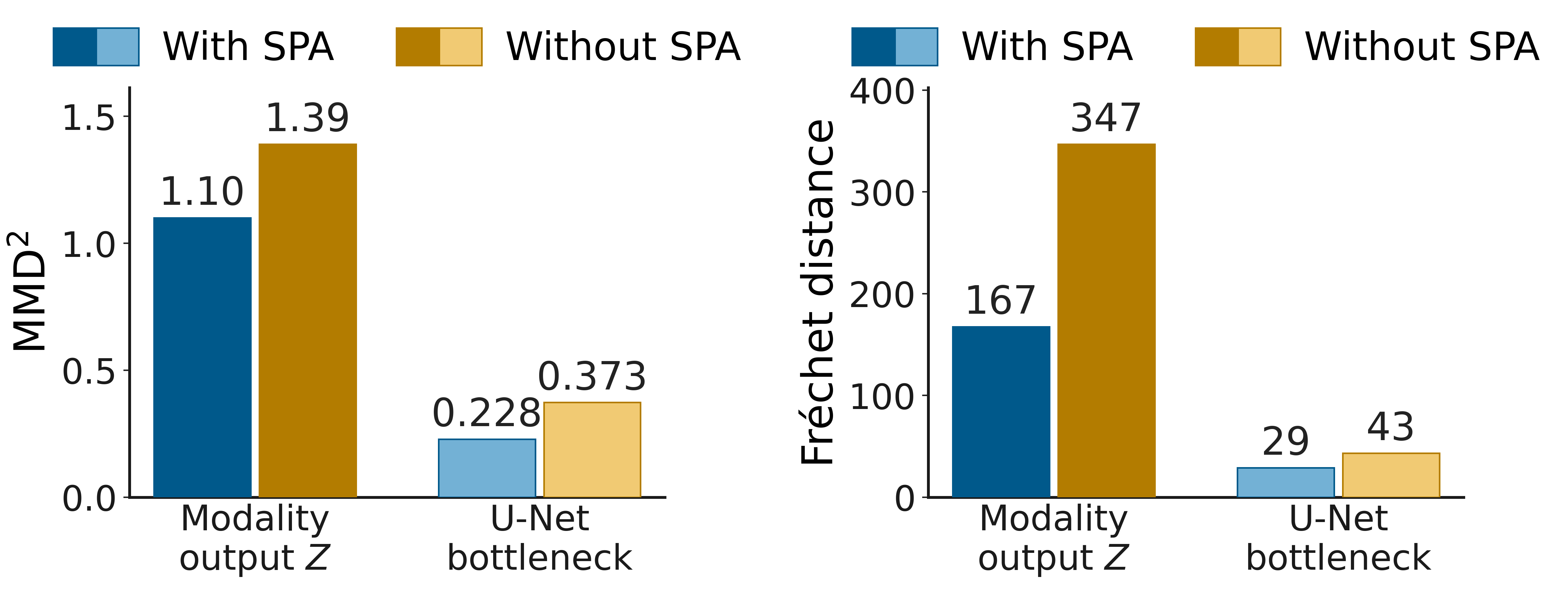}
\caption{Cross-client modality gap with and without SPA, measured at the modality-alignment output $Z^{k}_{i}$ and the U-Net bottleneck using MMD$^2$ (left) and Fr\'echet distance (right). Each bar is the mean over all client pairs; lower values indicates better alignment.}
\label{fig:spa_mmd}
\end{figure}

Figure~\ref{fig:spa_mmd} reports MMD$^2$ and Fr\'{e}chet distances between client foreground feature distributions on the patient-aligned evaluation set, with and without SPA. Enabling SPA consistently reduces the cross-client modality gap at both network depths ($Z^{k}_{i}$ and at the U-Net bottleneck) and under both metrics. At the modality-alignment output 
$Z^k_i$, SPA reduces mean pairwise MMD$^2$ from 1.39 to 1.10, a 21\% reduction, and Fr\'{e}chet distance from 346.8 to 167.4, a 52\% reduction. At the U-Net bottleneck, SPA reduces MMD$^2$ from 0.37 to 0.23 (38\% reduction) and Fr\'{e}chet distance from 43.4 to 28.9 (33\% reduction). The fact that alignment improves at both depths confirms that SPA regularizes not only the immediate modality-alignment output but propagates alignment benefits through the shared backbone, producing a more consistent fused representation at every level of the network consumed by the segmentation head. Because patient content is held constant across clients in the 
patient-aligned set, these reductions are attributable exclusively to the spectral alignment objective rather than sampling variability, providing direct causal evidence that SPA measurably reduces the distribution shift introduced by heterogeneous multimodal fusion in a federated setting.

\section{Discussion}
\label{sec:discussion}

%\textbf{Summary.} 
MOSAIC addresses the joint challenges of scarce annotations, private siloed data, and client-specific missing modalities by coupling a count-based modality-alignment module with a spectral prototype alignment loss and a dedicated federated refinement network. This two-phase design allows clients to train collaboratively from image-level labels alone, without raw data leaving any site and without assuming a fixed modality schema.

%\textbf{Image-level supervision suffices under missing modalities.} 
The central finding is that image-level supervision (cheapest form of annotation) is sufficient to train an accurate federated tumor segmentation model even when clients hold disjoint modality subsets. Across all three datasets, MOSAIC outperforms methods relying on substantially richer bounding-box and point annotations, narrowing the gap to fully supervised methods to under 0.05 Dice on FeTS2022, 0.03 on BraTS-MEN, and 0.04 on BraTS-SSA. Phase~1 pseudo-labels already outperform all competing weakly supervised methods, confirming that modality-robust CAM generation is the primary driver; Phase~2 adds consistent gains of 0.04, 0.02, and 0.03 Dice respectively by correcting CAM mislocalization and suppressing false positives, the two failure modes most directly amplified by missing modalities.

%\textbf{Federated multimodal fusion can exceed centralized fusion.} 
MOSAIC outperforms a centralized image-level model trained on pooled 
data with the complete modality set by 0.02 Dice on FeTS2022 and matches it within 0.01 Dice on BraTS-MEN. We attribute this to two mechanisms: the client-specific alignment module enforces site-specific fusion preprocessing adapted to each client's available channels, and the federated setting introduces implicit regularization through client-specific adaptation that avoids the representation confusion arising when a centralized model must learn from concatenated incomplete inputs. This finding suggests that privacy-preserving federated fusion with client-specific adapters is not merely a constrained approximation of centralized fusion but can be a qualitatively superior strategy when client distributions are heterogeneous.

%\textbf{Why frequency-domain alignment outperforms spatial alternatives.} 
The alignment comparison (Table~\ref{tab:alignment_obj}) reveals a deeper principle: first- and second-order spatial statistics are dominated by global intensity level and contrast, which differ across MRI sequences but do not capture modality-specific textural signatures. Histogram-KL, which matches the full marginal intensity distribution, is the closest alternative at 0.78 Dice, yet remains insensitive to the spatial frequency structure that encodes these signatures. SPA, by summarizing radial band energies in the 2-D Fourier domain, directly targets low, mid, and high-frequency patterns corresponding to global gradients, tissue-level textures, and fine structural detail, forcing the fused representation toward a globally consistent spectral profile regardless of which modalities contributed. This result suggests that frequency-domain prototype alignment will likely outperform spatial-statistic alternatives in any federated multimodal fusion task where modality-specific signatures manifest as distinct frequency patterns.

%\textbf{Cross-dataset generalisation as evidence of a general principle.}
The consistent superiority of SPA over FedNorm+, the only competing method designed for missing modalities, across all three structurally distinct modality configurations reveals a broader principle: aligning the spectral energy of the fused representation, which is invariant to which channels contributed, is more robust than normalizing modality-specific pathways that require knowing which modalities are present. This distinction may generalize beyond brain tumor MRI to any federated multimodal fusion setting with heterogeneous acquisition protocols, including cross-institutional pathology, federated EHR-imaging fusion, and multi-sensor clinical prediction.

%\textbf{Open federation and trustworthy uncertainty.} 
Conditioning on channel count rather than modality identity means the federation is not tied to a fixed institution list. Dynamic new client addition (Table~\ref{tab:newclient}) confirms that previously unseen sites reach within 0.01--0.04 Dice without retraining. Uncertainty analysis further shows that $\mathrm{Dice}_{\mathrm{unc}} < 0.035$ across all datasets, with uncertainty concentrated at tumor boundaries and correctly tracking the information content of available fused modalities, thus confirming calibrated rather than overconfident predictions under heterogeneous fusion inputs. Additionally, SPA shares only 137 floating-point values (548 bytes) per client per round, and the band-energy mapping is non-invertible by construction.

%\textbf{Privacy scope and future hardening.} 
% However, this does not constitute a formal differential privacy guarantee: an adversary with many rounds of shared statistics could infer distributional properties of client data. For deployments requiring formal guarantees, MOSAIC's compact statistic sharing is compatible with differential privacy~\cite{mcmahan2017learning} or secure aggregation~\cite{bonawitz2017practical} with minimal overhead, and this remains an important direction for trustworthy federated multimodal fusion.

%\textbf{Limitations and future work.} 
The current method is restricted to binary tumor segmentation; extending to multi-class sub-region segmentation is challenging because image-level labels become ambiguous when multiple nested classes coexist in one slice. Multi-label supervision or coarse spatial annotations could address this. The 2D slice-based implementation limits volumetric context; a fully 3D formulation would reduce boundary errors but requires stronger pseudo-label refinement to handle noisier 3D CAMs. Performance remains bounded by CAM quality, and more expressive weak localization, including foundation model-derived pseudo-labels~\cite{chen2024_wsss} or text-guided localization from radiology reports, could raise the accuracy ceiling. Finally, while validated on multi-parametric MRI, the frequency-domain alignment principle is modality-agnostic and should be evaluated in genuinely cross-modal federated settings combining MRI with PET, CT, or clinical time-series.

\section{Conclusions}
\label{sec:conclusion}
We propose MOSAIC, the first modality-agnostic federated framework for 
image-level weakly supervised tumor segmentation under client-specific 
missing modalities. A count-based modality-alignment module, spectral 
prototype alignment loss, and dedicated federated refinement network 
together enable collaborative training from image-level labels alone, 
without raw data leaving any site or modality bookkeeping. Across three 
multi-institutional brain tumor benchmarks, MOSAIC consistently 
outperforms all image-, box-, and point-supervised federated baselines, 
reaches 0.84 Dice on FeTS2022 using only image-level labels, and 
surpasses a centralized model with access to complete modality sets, 
demonstrating that privacy-preserving federated fusion can exceed naive 
data centralization. Dynamic new client addition enables previously 
unseen institutions to join within 0.01--0.04 Dice without retraining. 
These results establish frequency-domain spectral alignment as a 
general strategy for trustworthy multimodal fusion under heterogeneous 
and incomplete inputs. Future work will extend MOSAIC to multi-class 
segmentation, 3D formulations, and cross-modal federated settings.

\section*{Data availability and Ethics statement}
The datasets used in this study are publicly available through their respective repositories. The FeTS 2022 dataset is available via the \href{https://www.synapse.org/Synapse:syn28546456/wiki/633440}{FeTS 2022 Challenge repository on Synapse}, the BraTS 2023 intracranial meningioma dataset via the \href{https://www.synapse.org/Synapse:syn51514106}{BraTS 2023 Meningioma Challenge repository on Synapse}, and the BraTS-Africa (SSA) dataset via \href{https://www.kaggle.com/datasets/aiocta/brats2023-ssa-training-dataset}{BraTS 2023 SSA dataset repository on Kaggle}. All datasets were collected under appropriate institutional ethical oversight. FeTS 2022 was released following institutional IRB approval with informed consent obtained from participants; the BraTS 2023 meningioma dataset was collected under IRB approval with informed-consent waivers and anonymization; and the BraTS-Africa study was approved by the relevant institutional ethics boards, with informed-consent requirements waived.
This study involved secondary
analysis of existing de-identified research data and did
not involve direct interaction with human participants.
Accordingly, no additional institutional ethical approval
was required for this retrospective secondary analysis
under the applicable data use agreements.

\section*{CRediT authorship contribution statement}
\textbf{Tarun Kumar Garg:} Conceptualization, Methodology, Software, Validation,
Formal analysis, Investigation, Visualization, Writing - original draft.
\textbf{Vaanathi Sundaresan:} Conceptualization, Methodology, Data curation,
Funding acquisition, Project administration, Supervision, Writing - review and
editing.

\section*{Declaration of Generative AI and AI-assisted technologies in the writing process}
This manuscript was written entirely by the authors
without the use of large language models (LLMs) or artificial intelligence tools. All content, including analysis, interpretations, and conclusions, is solely the product of the authors’ research and intellectual effort.

\section*{Declaration of competing interest}
The authors declare that they have no known competing financial interests or personal relationships that could have appeared to influence the work reported in
this paper.

\section*{Acknowledgments}
This work was supported by DBT/Wellcome Trust
India Alliance Fellowship [IA/E/22/1/506763]. This
work was also supported in part by a grant from the
Council of Scientific $\&$ Industrial Research (CSIR) under its ASPIRE (Women Scientist Scheme) program
[25WS(013)/2023-24/EMR-II/ASPIRE] and in part by
Start-up Research Grant [SRG/2023/001406] from the
Science and Engineering Research Board, India. VS
is also supported by Pratiksha Trust, Bangalore, India [FG/PTCH-23-1004] and the Seed Research Grant
[IE/RERE-22-0583] from the Indian Institute of Science, India.
\bibliographystyle{unsrtnat}
\bibliography{ref}

@article{chen2024_wsss,
  title={Weakly-supervised semantic segmentation with image-level labels: from traditional models to foundation models},
  author={Chen, Zhaozheng and Sun, Qianru},
  journal={ACM Computing Surveys},
  volume={57},
  number={5},
  pages={1--29},
  year={2025},
  publisher={ACM New York, NY}
}

@article{sheller2020federated,
  title={Federated learning in medicine: facilitating multi-institutional collaborations without sharing patient data},
  author={Sheller, Micah J and Edwards, Brandon and Reina, G Anthony and Martin, Jason and Pati, Sarthak and Kotrotsou, Aikaterini and Milchenko, Mikhail and Xu, Weilin and Marcus, Daniel and Colen, Rivka R and others},
  journal={Scientific reports},
  volume={10},
  number={1},
  pages={12598},
  year={2020},
  publisher={Nature Publishing Group UK London}
}

@article{feddm2023tmi,
  title={FedDM: Federated weakly supervised segmentation via annotation calibration and gradient de-conflicting},
  author={Zhu, Meilu and Chen, Zhen and Yuan, Yixuan},
  journal={IEEE Transactions on Medical Imaging},
  volume={42},
  number={6},
  pages={1632--1643},
  year={2023},
  publisher={IEEE}
}

@inproceedings{selvaraju2017_gradcam,
  title={Grad-cam: Visual explanations from deep networks via gradient-based localization},
  author={Selvaraju, Ramprasaath R and Cogswell, Michael and Das, Abhishek and Vedantam, Ramakrishna and Parikh, Devi and Batra, Dhruv},
  booktitle={Proceedings of the IEEE international conference on computer vision},
  pages={618--626},
  year={2017}
}

@article{bernecker2022_fednorm,
  title={Fednorm: Modality-based normalization in federated learning for multi-modal liver segmentation},
  author={Bernecker, Tobias and Peters, Annette and Schlett, Christopher L and Bamberg, Fabian and Theis, Fabian and Rueckert, Daniel and Wei{\ss}, Jakob and Albarqouni, Shadi},
  journal={arXiv preprint arXiv:2205.11096},
  year={2022}
}

@inproceedings{peng2024_fedmm,
  title={Fedmm: Federated multi-modal learning with modality heterogeneity in computational pathology},
  author={Peng, Yuanzhe and Bian, Jieming and Xu, Jie},
  booktitle={ICASSP 2024-2024 IEEE International Conference on Acoustics, Speech and Signal Processing (ICASSP)},
  pages={1696--1700},
  year={2024},
  organization={IEEE}
}

@article{liu2025_fedmepd,
  title={Federated modality-specific encoders and partially personalized fusion decoder for multimodal brain tumor segmentation},
  author={Liu, Hong and Wei, Dong and Dai, Qian and Wu, Xian and Zheng, Yefeng and Wang, Liansheng},
  journal={Medical Image Analysis},
  pages={103759},
  year={2025},
  publisher={Elsevier}
}

@inproceedings{dhamale2025_interclass,
  title={Inter-class separability loss for weakly supervised mutually exclusive multiclass segmentation of brain tumor lesions},
  author={Dhamale, Vivek and Sundaresan, Vaanathi},
  booktitle={International Conference on Medical Image Computing and Computer-Assisted Intervention},
  pages={247--257},
  year={2025},
  organization={Springer}
}

@inproceedings{score_cam,
  title={Score-CAM: Score-weighted visual explanations for convolutional neural networks},
  author={Wang, Haofan and Wang, Zifan and Du, Mengnan and Yang, Fan and Zhang, Zijian and Ding, Sirui and Mardziel, Piotr and Hu, Xia},
  booktitle={2020 IEEE/CVF conference on computer vision and pattern recognition workshops (CVPRW)},
  pages={111--119},
  year={2020},
  organization={IEEE}
}

@article{jiang2021layercam,
  title={Layercam: Exploring hierarchical class activation maps for localization},
  author={Jiang, Peng-Tao and Zhang, Chang-Bin and Hou, Qibin and Cheng, Ming-Ming and Wei, Yunchao},
  journal={IEEE transactions on image processing},
  volume={30},
  pages={5875--5888},
  year={2021},
  publisher={IEEE}
}

@inproceedings{ame_cam,
  title={Ame-cam: Attentive multiple-exit cam for weakly supervised segmentation on mri brain tumor},
  author={Chen, Yu-Jen and Hu, Xinrong and Shi, Yiyu and Ho, Tsung-Yi},
  booktitle={International Conference on Medical Image Computing and Computer-Assisted Intervention},
  pages={173--182},
  year={2023},
  organization={Springer}
}

@article{reina2021openfl_fets_data,
  title={Openfl: An open-source framework for federated learning},
  author={Reina, G Anthony and Gruzdev, Alexey and Foley, Patrick and Perepelkina, Olga and Sharma, Mansi and Davidyuk, Igor and Trushkin, Ilya and Radionov, Maksim and Mokrov, Aleksandr and Agapov, Dmitry and others},
  journal={arXiv preprint arXiv:2105.06413},
  year={2021}
}

@article{wang2025fedfat,
  title={FedFAT: Frequency adpative interpolation for federated domain generalization on heterogeneous medical images},
  author={Wang, Donghao and Cui, Yingchun and Li, Mingyang and Xi, Heran and Zhu, Jinghua},
  journal={Pattern Recognition},
  pages={112459},
  year={2025},
  publisher={Elsevier}
}

@inproceedings{ronneberger2015u,
  title={U-net: Convolutional networks for biomedical image segmentation},
  author={Ronneberger, Olaf and Fischer, Philipp and Brox, Thomas},
  booktitle={International Conference on Medical image computing and computer-assisted intervention},
  pages={234--241},
  year={2015},
  organization={Springer}
}

@article{li2020federated,
  title={Federated optimization in heterogeneous networks},
  author={Li, Tian and Sahu, Anit Kumar and Zaheer, Manzil and Sanjabi, Maziar and Talwalkar, Ameet and Smith, Virginia},
  journal={Proceedings of Machine learning and systems},
  volume={2},
  pages={429--450},
  year={2020}
}

@inproceedings{mcmahan2017communication,
  title={Communication-efficient learning of deep networks from decentralized data},
  author={McMahan, Brendan and Moore, Eider and Ramage, Daniel and Hampson, Seth and y Arcas, Blaise Aguera},
  booktitle={Artificial intelligence and statistics},
  pages={1273--1282},
  year={2017},
  organization={Pmlr}
}

@article{tarvainen2017mean,
  title={Mean teachers are better role models: Weight-averaged consistency targets improve semi-supervised deep learning results},
  author={Tarvainen, Antti and Valpola, Harri},
  journal={Advances in neural information processing systems},
  volume={30},
  year={2017}
}

@inproceedings{tan2022fedproto,
  title={Fedproto: Federated prototype learning across heterogeneous clients},
  author={Tan, Yue and Long, Guodong and Liu, Lu and Zhou, Tianyi and Lu, Qinghua and Jiang, Jing and Zhang, Chengqi},
  booktitle={Proceedings of the AAAI conference on artificial intelligence},
  volume={36},
  number={8},
  pages={8432--8440},
  year={2022}
}

@inproceedings{abraham2019focaltversky,
  title={A novel focal tversky loss function with improved attention u-net for lesion segmentation},
  author={Abraham, Nabila and Khan, Naimul Mefraz},
  booktitle={2019 IEEE 16th international symposium on biomedical imaging (ISBI 2019)},
  pages={683--687},
  year={2019},
  organization={IEEE}
}

@article{obukhov2019gatedcrf,
  title={Gated CRF loss for weakly supervised semantic image segmentation},
  author={Obukhov, Anton and Georgoulis, Stamatios and Dai, Dengxin and Van Gool, Luc},
  journal={arXiv preprint arXiv:1906.04651},
  year={2019}
}

@inproceedings{lin2016scribblesup,
  title={Scribblesup: Scribble-supervised convolutional networks for semantic segmentation},
  author={Lin, Di and Dai, Jifeng and Jia, Jiaya and He, Kaiming and Sun, Jian},
  booktitle={Proceedings of the IEEE conference on computer vision and pattern recognition},
  pages={3159--3167},
  year={2016}
}

@inproceedings{wang2020seam,
  title={Self-supervised equivariant attention mechanism for weakly supervised semantic segmentation},
  author={Wang, Yude and Zhang, Jie and Kan, Meina and Shan, Shiguang and Chen, Xilin},
  booktitle={2020 IEEE/CVF conference on computer vision and pattern recognition (CVPR)},
  pages={12272--12281},
  year={2020},
  organization={IEEE}
}

@inproceedings{li2021moon,
  title={Model-contrastive federated learning},
  author={Li, Qinbin and He, Bingsheng and Song, Dawn},
  booktitle={2021 IEEE/CVF Conference on Computer Vision and Pattern Recognition (CVPR)},
  pages={10708--10717},
  year={2021},
  organization={IEEE}
}

@inproceedings{lin2023fedicra,
  title={Unifying and personalizing weakly-supervised federated medical image segmentation via adaptive representation and aggregation},
  author={Lin, Li and Wu, Jiewei and Liu, Yixiang and Wong, Kenneth KY and Tang, Xiaoying},
  booktitle={International Workshop on Machine Learning in Medical Imaging},
  pages={196--206},
  year={2023},
  organization={Springer}
}

@article{lin2025fedlppa,
  title={FedLPPA: Learning personalized prompt and aggregation for federated weakly-supervised medical image segmentation},
  author={Lin, Li and Liu, Yixiang and Wu, Jiewei and Cheng, Pujin and Cai, Zhiyuan and Wong, Kenneth KY and Tang, Xiaoying},
  journal={IEEE Transactions on Medical Imaging},
  volume={44},
  number={3},
  pages={1127--1139},
  year={2024},
  publisher={IEEE}
}

@inproceedings{ding2021rfnet,
  title={RFNet: Region-aware fusion network for incomplete multi-modal brain tumor segmentation},
  author={Ding, Yuhang and Yu, Xin and Yang, Yi},
  booktitle={2021 IEEE/CVF International Conference on Computer Vision (ICCV)},
  pages={3955--3964},
  year={2021},
  organization={IEEE}
}

@inproceedings{hu2020kdnet,
  title={Knowledge distillation from multi-modal to mono-modal segmentation networks},
  author={Hu, Minhao and Maillard, Matthis and Zhang, Ya and Ciceri, Tommaso and La Barbera, Giammarco and Bloch, Isabelle and Gori, Pietro},
  booktitle={International Conference on Medical Image Computing and Computer-Assisted Intervention},
  pages={772--781},
  year={2020},
  organization={Springer}
}

@inproceedings{yang2020fda,
  title={{FDA}: Fourier domain adaptation for semantic segmentation},
  author={Yang, Yanchao and Soatto, Stefano},
  booktitle={Proceedings of the IEEE/CVF Conference on Computer Vision and Pattern Recognition (CVPR)},
  pages={4085--4095},
  year={2020}
}

@inproceedings{jiang2022harmofl,
  title={Harmofl: Harmonizing local and global drifts in federated learning on heterogeneous medical images},
  author={Jiang, Meirui and Wang, Zirui and Dou, Qi},
  booktitle={Proceedings of the AAAI conference on artificial intelligence},
  volume={36},
  number={1},
  pages={1087--1095},
  year={2022}
}

@inproceedings{liu2021feddg,
  title={{FedDG}: Federated domain generalization on medical image segmentation via episodic learning in continuous frequency space},
  author={Liu, Quande and Chen, Cheng and Qin, Jing and Dou, Qi and Heng, Pheng-Ann},
  booktitle={Proceedings of the IEEE/CVF Conference on Computer Vision and Pattern Recognition (CVPR)},
  pages={1013--1023},
  year={2021}
}

@inproceedings{pan2025fdgpolyp,
  title={Frequency-based federated domain generalization for polyp segmentation},
  author={Pan, Hongyi and Jha, Debesh and Biswas, Koushik and Bagci, Ulas},
  booktitle={ICASSP 2025-2025 IEEE International Conference on Acoustics, Speech and Signal Processing (ICASSP)},
  pages={1--5},
  year={2025},
  organization={IEEE}
}

@article{fets2022,
  title={The Federated Tumor Segmentation (FeTS) challenge},
  author={Pati, Sarthak and Baid, Ujjwal and Edwards, Brandon and Sheller, Micah J and Foley, Patrick and Reina, G Anthony and Thakur, Siddhesh and Sako, Chiharu and Bilello, Michel and Davatzikos, Christos and others},
  journal={arXiv preprint arXiv:2105.05874},
  year={2021}
}

@article{madni2025fl,
  title={{FL-W3S}: Cross-domain federated learning for weakly supervised semantic segmentation of white blood cells},
  author={Madni, Hussain Ahmad and Umer, Rao Muhammad and Zottin, Silvia and Marr, Carsten and Foresti, Gian Luca},
  journal={International Journal of Medical Informatics},
  volume={195},
  pages={105806},
  year={2025},
  publisher={Elsevier},
  doi={10.1016/j.ijmedinf.2025.105806}
}

@inproceedings{ahn2019irnet,
  title={Weakly supervised learning of instance segmentation with inter-pixel relations},
  author={Ahn, Jiwoon and Cho, Sunghyun and Kwak, Suha},
  booktitle={2019 IEEE/CVF Conference on Computer Vision and Pattern Recognition (CVPR)},
  pages={2204--2213},
  year={2019},
  organization={IEEE}
}

@article{fallah2020personalized,
  title={Personalized federated learning: A meta-learning approach},
  author={Fallah, Alireza and Mokhtari, Aryan and Ozdaglar, Asuman},
  journal={arXiv preprint arXiv:2002.07948},
  year={2020}
}

@inproceedings{zhang2022mmformer,
  title={mmformer: Multimodal medical transformer for incomplete multimodal learning of brain tumor segmentation},
  author={Zhang, Yao and He, Nanjun and Yang, Jiawei and Li, Yuexiang and Wei, Dong and Huang, Yawen and Zhang, Yang and He, Zhiqiang and Zheng, Yefeng},
  booktitle={International conference on medical image computing and computer-assisted intervention},
  pages={107--117},
  year={2022},
  organization={Springer}
}

@inproceedings{xu2022multi,
  title={Multi-class token transformer for weakly supervised semantic segmentation},
  author={Xu, Lian and Ouyang, Wanli and Bennamoun, Mohammed and Boussaid, Farid and Xu, Dan},
  booktitle={2022 IEEE/CVF Conference on Computer Vision and Pattern Recognition (CVPR)},
  pages={4300--4309},
  year={2022},
  organization={IEEE}
}

@article{qin2023portable,
  title={Portable skin lesion segmentation system with accurate lesion localization based on weakly supervised learning},
  author={Qin, Hai and Deng, Zhanjin and Shu, Liye and Yin, Yi and Li, Jintao and Zhou, Li and Zeng, Hui and Liang, Qiaokang},
  journal={Electronics},
  volume={12},
  number={17},
  pages={3732},
  year={2023},
  publisher={MDPI}
}

@inproceedings{xie2023exploring,
  title={Exploring and exploiting uncertainty for incomplete multi-view classification},
  author={Xie, Mengyao and Han, Zongbo and Zhang, Changqing and Bai, Yichen and Hu, Qinghua},
  booktitle={2023 IEEE/CVF Conference on Computer Vision and Pattern Recognition (CVPR)},
  pages={19873--19882},
  year={2023},
  organization={IEEE}
}

@article{mcmahan2017learning,
  title={Learning differentially private recurrent language models},
  author={McMahan, H Brendan and Ramage, Daniel and Talwar, Kunal and Zhang, Li},
  journal={arXiv preprint arXiv:1710.06963},
  year={2017}
}

@inproceedings{bonawitz2017practical,
  title={Practical secure aggregation for privacy-preserving machine learning},
  author={Bonawitz, Keith and Ivanov, Vladimir and Kreuter, Ben and Marcedone, Antonio and McMahan, H Brendan and Patel, Sarvar and Ramage, Daniel and Segal, Aaron and Seth, Karn},
  booktitle={proceedings of the 2017 ACM SIGSAC Conference on Computer and Communications Security},
  pages={1175--1191},
  year={2017}
}

@inproceedings{hayat2022medfuse,
  title={MedFuse: Multi-modal fusion with clinical time-series data and chest X-ray images},
  author={Hayat, Nasir and Geras, Krzysztof J and Shamout, Farah E},
  booktitle={Machine Learning for Healthcare Conference},
  pages={479--503},
  year={2022},
  organization={PMLR}
}

@inproceedings{gatys2016image,
  title={Image style transfer using convolutional neural networks},
  author={Gatys, Leon A and Ecker, Alexander S and Bethge, Matthias},
  booktitle={Proceedings of the IEEE conference on computer vision and pattern recognition},
  pages={2414--2423},
  year={2016}
}

@article{labella2023bratsmen,
  title={The asnr-miccai brain tumor segmentation (brats) challenge 2023: Intracranial meningioma},
  author={LaBella, Dominic and Adewole, Maruf and Alonso-Basanta, Michelle and Altes, Talissa and Anwar, Syed Muhammad and Baid, Ujjwal and Bergquist, Timothy and Bhalerao, Radhika and Chen, Sully and Chung, Verena and others},
  journal={arXiv preprint arXiv:2305.07642},
  year={2023}
}

@article{adewole2023brain,
  title={The brain tumor segmentation (brats) challenge 2023: Glioma segmentation in sub-saharan africa patient population (brats-africa)},
  author={Adewole, Maruf and Rudie, Jeffrey D and Gbdamosi, Anu and Toyobo, Oluyemisi and Raymond, Confidence and Zhang, Dong and Omidiji, Olubukola and Akinola, Rachel and Suwaid, Mohammad Abba and Emegoakor, Adaobi and others},
  journal={ArXiv},
  pages={arXiv--2305},
  year={2023}
}

@inproceedings{rong2023boundary,
  title={Boundary-enhanced co-training for weakly supervised semantic segmentation},
  author={Rong, Shenghai and Tu, Bohai and Wang, Zilei and Li, Junjie},
  booktitle={Proceedings of the IEEE/CVF conference on computer vision and pattern recognition},
  pages={19574--19584},
  year={2023}
}

@article{baid2021rsna,
  title={The rsna-asnr-miccai brats 2021 benchmark on brain tumor segmentation and radiogenomic classification},
  author={Baid, Ujjwal and Ghodasara, Satyam and Mohan, Suyash and Bilello, Michel and Calabrese, Evan and Colak, Errol and Farahani, Keyvan and Kalpathy-Cramer, Jayashree and Kitamura, Felipe C and Pati, Sarthak and others},
  journal={arXiv preprint arXiv:2107.02314},
  year={2021}
}

@article{karargyris2023federated,
  title={Federated benchmarking of medical artificial intelligence with MedPerf},
  author={Karargyris, Alexandros and Umeton, Renato and Sheller, Micah J and Aristizabal, Alejandro and George, Johnu and Wuest, Anna and Pati, Sarthak and Kassem, Hasan and Zenk, Maximilian and Baid, Ujjwal and others},
  journal={Nature machine intelligence},
  volume={5},
  number={7},
  pages={799--810},
  year={2023},
  publisher={Nature Publishing Group UK London}
}

@inproceedings{tang2018regularized,
  title={On regularized losses for weakly-supervised cnn segmentation},
  author={Tang, Meng and Perazzi, Federico and Djelouah, Abdelaziz and Ben Ayed, Ismail and Schroers, Christopher and Boykov, Yuri},
  booktitle={Proceedings of the European conference on computer vision (ECCV)},
  pages={507--522},
  year={2018}
}

@inproceedings{
li2021fedbn,
title={Fed{BN}: Federated Learning on Non-{IID} Features via Local Batch Normalization},
author={Xiaoxiao Li and Meirui JIANG and Xiaofei Zhang and Michael Kamp and Qi Dou},
booktitle={International Conference on Learning Representations},
year={2021},
url={https://openreview.net/forum?id=6YEQUn0QICG}
}

\end{document}